\documentclass[journal]{IEEEtran}

\usepackage{amsthm}
\usepackage{amsmath,amsfonts}
\usepackage{algorithm}
\usepackage{algpseudocode}
\usepackage{array}
\usepackage[caption=false,font=normalsize,labelfont=sf,textfont=sf]{subfig}
\usepackage{textcomp}
\usepackage{stfloats}
\usepackage{float}
\usepackage{url}
\usepackage{verbatim}
\usepackage{graphicx}
\graphicspath{{figs/}}
\usepackage{import}
\usepackage{xcolor} 

\usepackage{transparent}
\usepackage{calc}
\usepackage{cite}
\usepackage{psfrag}
\usepackage{enumerate}
\usepackage{amssymb}
\usepackage{booktabs}
\usepackage{multirow}
\usepackage{diagbox}
\usepackage{pifont}
\def\BibTeX{{\rm B\kern-.05em{\sc i\kern-.025em b}\kern-.08em
		T\kern-.1667em\lower.7ex\hbox{E}\kern-.125emX}}

\newtheorem{remark}{Remark}
\newtheorem{proposition}{Proposition}

{
}

\newcommand{\putFig}[3]{
        \begin{figure}[htbp!]
		\centering
		\includegraphics[width=#3in]{figs/#1}
                \caption{#2}
                \label{fig:#1}
        \end{figure} }

\begin{document}
	\bibliographystyle{IEEEtran}
	\title{LOLLA: Deep Reinforcement Learning for Closed-Loop\\ Link Adaptation Towards a GPU-Accelerated AI-RAN}
	
	\author{Rui~Wang,~\IEEEmembership{Member,~IEEE,}
		Linchao~Zhang,~\IEEEmembership{Member,~IEEE,}
		Qiang~Liu,~\IEEEmembership{Senior~Member,~IEEE,}
		and~Kun~Yang,~\IEEEmembership{Fellow,~IEEE}
		\thanks{Rui Wang, Linchao Zhang, and Qiang Liu are with the School of Information and Communication Engineering, University of Electronic Science and Technology of China, Chengdu 611731, China, and also with the Yangtze Delta Region Institute (Quzhou), University of Electronic Science and Technology of China, Quzhou 324003, China (e-mail: happytank888@csj.uestc.edu.cn; zhanglinchao@csj.uestc.edu.cn; liuqiang@uestc.edu.cn).}
		\thanks{Kun Yang is with the State Key Laboratory of Novel Software Technology, Nanjing University, Nanjing 210008, China, with the Institute of Intelligent Networks and Communications (NINE) and School of Intelligent Software and Engineering, Nanjing University (Suzhou Campus), Suzhou 215163, China, and also with the School of Information and Communication Engineering, University of Electronic Science and Technology of China, Chengdu 611731, China (e-mail: kunyang@nju.edu.cn).}}

	\maketitle

	\begin{abstract} 
		Outer-loop link adaptation (OLLA) is widely deployed in 5G NR to track channel variations, yet its reliance on first-order, single-bit feedback degrades performance significantly under high-mobility and fast-varying channels. This paper presents \textit{LOLLA} (\textbf{L}earned \textbf{O}uter-\textbf{L}oop \textbf{L}ink \textbf{A}daptation), a deep reinforcement learning framework that replaces the conventional OLLA staircase with a learned, continuous SINR offset conditioned on rich PHY/MAC telemetry inaccessible to OLLA. The offset modulates the SINR-to-MCS lookup table, preserving 3GPP-compliant MCS selection and provably subsuming the conventional OLLA update rule. A Proximal Policy Optimization (PPO) policy trained under a Lagrangian block error rate (BLER) constraint automatically enforces tunable reliability targets from 1\% to 15\% without manual penalty calibration. The framework is realized as the first closed-loop AI-native control dApp on a GPU-accelerated 5G NR stack, achieving end-to-end control latencies under 500\,$\mu$s. Evaluations under 3GPP TDL channel models demonstrate 15\% to 92\% throughput gains over OLLA across Doppler frequencies up to 400\,Hz, while attaining a Pareto frontier that strictly dominates OLLA across all evaluated reliability targets. The learned policy generalizes to unseen channel models and scales to eight concurrent UEs under shared-resource scheduling. In the uplink formulation, the gNB directly observes decoding outcomes, enabling simulation-to-deployment parity.
	\end{abstract}
	

	\section{Introduction}
	\label{sec:introduction}
	
	\IEEEPARstart{T}he transition towards 5G-Advanced and 6G Radio Access Networks (RANs) is driven by a paradigm shift from fixed-function hardware to software-defined, virtualized architectures running on programmable computing platforms~\cite{Jiang2021TheRT}. This evolution, exemplified by the O-RAN Alliance's disaggregated architecture~\cite{Polese:ICST:2023}, enables the deployment of intelligent controllers, including xApps for near-real-time and dApps for real-time control~\cite{DOro2022}, that can optimize network performance at unprecedented granularity. GPU-accelerated baseband processing~\cite{Kelkar2021NVIDIAAG} has emerged as a key enabler, providing the computational power for massive MIMO signal processing and the ability to expose fine-grained physical layer (PHY) telemetry at slot level~\cite{Villa2025ProgrammableAG}. In particular, dApps, lightweight applications co-located with the baseband processing unit, can access PHY/MAC telemetry and execute control actions at sub-millisecond timescales~\cite{LACAVA2025111342}, making them uniquely suited for latency-critical Radio Resource Management (RRM) functions such as link adaptation.
	
	Link adaptation, the process of selecting the optimal Modulation and Coding Scheme (MCS) for each transmission, is a critical RRM function that directly impacts spectral efficiency and link reliability~\cite{Goldsmith:ITC:1998}. In 5G NR, link adaptation employs a two-loop structure: an inner loop that maps a Channel Quality Indicator (CQI) to an MCS index via a static SINR-to-MCS Lookup Table (LUT), and an Outer Loop Link Adaptation (OLLA) mechanism that adjusts an SINR offset based on HARQ ACK/NACK feedback to maintain a target Block Error Rate (BLER)~\cite{BlanquezCasado2016eOLLAAE}. While computationally efficient, OLLA is inherently reactive: each update consumes only one bit of feedback per transmission, rendering it slow to adapt under fast-varying channels in high-mobility scenarios~\cite{3GPP_TS38901,3GPP_TS37985} or bursty interference environments~\cite{Martín:IA:2021}. Deep Reinforcement Learning (DRL) can address these limitations by learning non-linear control policies from high-dimensional PHY state~\cite{Luong:ICST:2019}, and several recent works report substantial throughput gains over OLLA in simulation~\cite{Saxena:ITWC:2022,Ye:TVT:2023,Mota:GW:2019}; however, bridging the gap to real-time deployment remains challenging due to sub-millisecond control budgets~\cite{LACAVA2025111342}, the risk of discarding the structured SINR-to-MCS relationship encoded in the LUT, and the overhead of transporting high-dimensional PHY telemetry to an external controller.
	
	In this paper, we present \textit{LOLLA} (\textbf{L}earned \textbf{O}uter-\textbf{L}oop \textbf{L}ink \textbf{A}daptation), a real-time, closed-loop DRL controller for uplink link adaptation on GPU-accelerated RANs. Our agent learns a continuous SINR offset, a residual correction~\cite{Johannink:ICRA:2019}, that modulates the legacy SINR-to-MCS LUT, ensuring safe, standards-compliant default behavior when the offset is near zero while enabling the policy to exploit rich PHY/MAC telemetry inaccessible to OLLA. A Proximal Policy Optimization (PPO)-Lagrangian BLER constraint~\cite{Schulman2017ProximalPO,Altman1999CMDP,Achiam2019BenchmarkingSE} replaces manual penalty tuning, enabling automatic adaptation to arbitrary reliability targets. We realize this within the E3 dApp architecture~\cite{Villa2025ProgrammableAG} via a dual-agent design with zero-copy shared-memory data transfer, achieving sub-500\,$\mu\text{s}$ latency. The uplink formulation ensures the gNB both controls the MCS via uplink grants~\cite{3GPP_TS38212} and directly observes decoding outcomes, enabling simulation-to-deployment interface parity.
	
	\subsection{Related Work}
	\label{subsec:related_work}
	
	The limitations of conventional OLLA have motivated enhanced heuristics such as eOLLA~\cite{BlanquezCasado2016eOLLAAE} with adaptive step sizes and NOLLA~\cite{Zhu:PIMRC:2023} with non-linear offset adjustments, as well as SALAD~\cite{SALAD2025}, which replaces the fixed OLLA step with online gradient descent and demonstrates over-the-air gains on a GPU-accelerated testbed. On the learning side, approaches range from multi-armed~\cite{Khosravani2017} and contextual bandits~\cite{Saxena_CMAB2019}, latent Thompson Sampling~\cite{Saxena:ITWC:2022}, and supervised online learning~\cite{MassiveMIMO_ODL2022}, to Q-learning~\cite{Mota:GW:2019}, DQN~\cite{Ye:TVT:2023}, branching DQN for joint MCS and MIMO-layer adaptation~\cite{An2024DRAGONAD}, TD3 with Bayesian optimization for joint device scheduling and link adaptation in URLLC~\cite{Gao2025URLLC}, graph attention networks for cross-scenario generalization~\cite{GenRL_RAN2025}, and decoupled DQN addressing inference latency~\cite{You2026DCDQN}. Table~\ref{tab:comparison} provides a structured comparison.
	
	On the systems side, the E3 framework~\cite{Villa2025ProgrammableAG} establishes dApps for GPU-native RAN programmability with sub-millisecond PHY/MAC access. Recent deployments, including X5G~\cite{Villa_X5G2025}, interference detection~\cite{Santhi2025}, spectrum classification~\cite{Olimpieri2025LibIQTR}, and sim-to-field transfer~\cite{Ford2025Sim2FieldED}, all target open-loop inference without closed-loop actuation. Although SALAD~\cite{SALAD2025} achieves closed-loop SINR estimation, it relies on model-based gradient descent rather than learned policies. Across these works, three gaps persist: most use discrete MCS actions that bypass the SINR-to-MCS LUT, rely on fixed penalties or heuristic step sizes for BLER control rather than formal constrained optimization~\cite{Altman1999CMDP}, and lack closed-loop integration with a production-grade GPU-accelerated RAN stack.
	
	\begin{table*}[!t]
		\caption{Qualitative comparison of link adaptation methods. Cont.\ = continuous, Disc.\ = discrete, CL = closed-loop, Sim = simulation, OTA = over-the-air, LUT = SINR-to-MCS lookup table, Q-learn.\ = Q-learning, LTS = latent Thompson sampling, SL = supervised learning, Br.\ = branching, opt.\ = optimization, GAT = graph attention network.}
		\label{tab:comparison}
		\centering
		\small
		\begin{tabular*}{\textwidth}{@{\extracolsep{\fill}}lrcll ccc c@{}}
			\toprule
			& & & & & \multicolumn{3}{c}{\textbf{System capabilities}} & \\
			\cmidrule(lr){6-8}
			\textbf{Method} & \textbf{Year} & \textbf{Type} & \textbf{Action} & \textbf{BLER control} & \textbf{Multi-UE$^\ddagger$} & \textbf{CL$^\S$} & \textbf{GPU$^\dagger$} & \textbf{Validation} \\
			\midrule
			eOLLA~\cite{BlanquezCasado2016eOLLAAE} & 2016 & Heuristic & Cont.\ offset $\rightarrow$ LUT & Adaptive step & \ding{55} & \ding{55} & \ding{55} & Sim \\
			Mota~et~al.~\cite{Mota:GW:2019} & 2019 & RL (Q-learn.) & Disc.\ MCS & Reward penalty & \ding{55} & \ding{55} & \ding{55} & Sim \\
			Saxena~et~al.~\cite{Saxena:ITWC:2022} & 2022 & Bandit (LTS) & Disc.\ MCS & Bayesian & \ding{55} & \ding{55} & \ding{55} & Sim \\
			Bobrov~et~al.~\cite{MassiveMIMO_ODL2022} & 2022 & SL (online) & Disc.\ MCS & Implicit & \ding{55} & \ding{55} & \ding{55} & Sim \\
			Ye~et~al.~\cite{Ye:TVT:2023} & 2023 & DRL (DQN) & Disc.\ MCS & Implicit & \ding{55} & \ding{55} & \ding{55} & Sim \\
			NOLLA~\cite{Zhu:PIMRC:2023} & 2023 & Heuristic & Cont.\ offset $\rightarrow$ LUT & Non-linear step & \ding{55} & \ding{55} & \ding{55} & Sim \\
			DRAGON~\cite{An2024DRAGONAD} & 2024 & DRL (Br.\ DQN) & Disc.\ MCS+layer & Implicit & \ding{55} & \ding{55} & \ding{55} & Sim \\
			GenRL-RAN~\cite{GenRL_RAN2025} & 2025 & DRL (GAT) & Disc.\ MCS & Implicit & \ding{55} & \ding{55} & \ding{55} & Sim \\
			SALAD~\cite{SALAD2025} & 2025 & Online opt. & Cont.\ SINR $\rightarrow$ LUT & Integral feedback & \ding{55} & $\checkmark$ & $\checkmark$ & Sim + OTA \\
			Gao~et~al.~\cite{Gao2025URLLC} & 2025 & DRL (TD3) & Disc.\ MCS & Reward penalty & \ding{55} & \ding{55} & \ding{55} & Sim \\
			DC-DQN-LA~\cite{You2026DCDQN} & 2026 & DRL (DQN) & Disc.\ MCS & Reward penalty & \ding{55} & $\checkmark$ & \ding{55} & Sim + OTA \\
			\midrule
			\textbf{LOLLA} & \textbf{2026} & \textbf{DRL (PPO)} & \textbf{Cont.\ offset $\rightarrow$ LUT} & \textbf{Lagrangian} & $\checkmark$ & $\checkmark$ & $\checkmark$ & \textbf{Sim} \\
			\bottomrule
			\multicolumn{9}{l}{\footnotesize $^\dagger$\,GPU-accelerated PHY with identical CUDA baseband kernels as production hardware~\cite{Villa2025ProgrammableAG}.} \\
			\multicolumn{9}{l}{\footnotesize $^\ddagger$\,Evaluated under shared-resource multi-UE scheduling ($K > 1$ UEs sharing PRB resources).} \\
			\multicolumn{9}{l}{\footnotesize $^\S$\,Closed-loop actuation: algorithm's decisions are executed by the PHY/MAC stack (not open-loop inference only).}
		\end{tabular*}
	\end{table*}
	
	\subsection{Contributions and Paper Organization}
	\label{subsec:contributions}
	
	The main contributions of this paper are as follows:
	
	\begin{enumerate}
		\item A residual DRL formulation~\cite{Johannink:ICRA:2019} in which the agent learns a bounded, continuous SINR offset that modulates a link-level calibrated SINR-to-MCS mapping~\cite{Lagen2020NewRP}, retaining this mapping as a structural prior and reducing the policy search to a one-dimensional correction over a 13-dimensional PHY/MAC observation space that captures channel dynamics inaccessible to conventional OLLA. Because the offset action space strictly contains the linear OLLA update rule as a special case, the learned policy class is at least as expressive as OLLA, providing a formal performance dominance guarantee.
		\item A Constrained MDP formulation~\cite{Altman1999CMDP} with PPO-Lagrangian~\cite{Achiam2019BenchmarkingSE} online dual-variable updates that replaces manual penalty tuning. The flat NACK penalty produces a non-uniform per-SNR BLER allocation that concentrates the error budget where throughput gains are largest, provably throughput-optimal under a convex, continuous MCS relaxation~\cite[\S5.5.3]{Boyd2004}. A single training framework supports arbitrary BLER targets from eMBB (${\approx}10\%$) to URLLC ($< 1\%$) by specifying $\varepsilon_\textup{target}$ without altering the policy or training procedure.
		\item The first closed-loop \emph{learning-based} control dApp on the NVIDIA ARC-OTA platform. Whereas the E3 framework of Villa~\emph{et al.}~\cite{Villa2025ProgrammableAG} demonstrates open-loop inference dApps (e.g., sensing and PHY-layer telemetry processing), LOLLA realizes a full closed control loop in which a learned policy actuates the per-slot MCS and observes the resulting decoding outcomes. We bridge DU-Low and DU-High through a dual E3-Agent architecture connected via shared-memory data transfer through the Aerial Data Lake, and drive inference through four progressively optimized backends (PyTorch, Triton gRPC, Triton in-process, and Triton C-API), achieving end-to-end control latencies under 500\,$\mu\text{s}$ at 30\,kHz numerology.
		\item To our knowledge, the first multi-UE closed-loop link adaptation system on a real-time GPU-accelerated RAN. Prior dApp deployments~\cite{Villa2025ProgrammableAG,Santhi2025,Olimpieri2025LibIQTR} and RL-based link adaptation methods~\cite{Saxena:ITWC:2022,Ye:TVT:2023,Mota:GW:2019,An2024DRAGONAD} are limited to single-UE operation. Scaling to $K$ concurrent UEs requires solving three tightly coupled challenges: (C1)~per-UE state isolation via RNTI-indexed observation and reward tracking, (C2)~MAC-RL interface isolation through a two-stage PRB-only scheduler that decouples RL MCS decisions from the production OLLA loop, and (C3)~exact per-UE Transport Block size propagation from the MAC scheduler through the E3 pipeline, eliminating the approximation error inherent in uniform PRB-splitting.
		\item Extensive evaluation under 3GPP TDL channel models across Doppler frequencies up to 400\,Hz, multiple BLER targets ($\varepsilon_\textup{target} \in \{1\%, 5\%, 9.1\%, 12\%\}$), multi-UE configurations ($K \in \{1, 2, 4, 8\}$), and four inference backends, demonstrating 15\% to 92\% throughput gains over OLLA while satisfying the prescribed BLER constraints, with the gain growing under faster channel dynamics and peaking at $f_d = 200$\,Hz.
	\end{enumerate}
	
	The remainder of this paper is organized as follows. Section~\ref{sec:system_model} presents the system model, formulates the link adaptation problem as a constrained MDP, and reviews conventional OLLA. Section~\ref{sec:learned_olla} details the proposed LOLLA framework, including the residual action design, PPO policy architecture, and Lagrangian BLER constraint. Section~\ref{sec:system_arch} describes the real-time system architecture and inference pipeline. Section~\ref{sec:sim_results} presents simulation results and ablation studies, and Section~\ref{sec:conclusions} concludes the paper.
	
	\textbf{Notation:}~Boldface lowercase and uppercase letters denote vectors ($\mathbf{o}$) and matrices ($\mathbf{H}$). $(\cdot)^\top$ is the transpose, $\hat{\cdot}$ denotes an estimate, $\mathbb{I}(\cdot)$ the indicator function. $\mathcal{N}(\mu, \sigma^2)$ and $\mathcal{CN}(\boldsymbol{\mu}, \mathbf{C})$ denote real and circularly-symmetric complex Gaussian distributions. $\mathbb{E}[\cdot]$ is expectation, $\|\cdot\|_\text{TV}$ total variation distance, and $[x]_a^b \triangleq \max(a, \min(b, x))$ the projection onto $[a,b]$. $\mathcal{M}$ the MCS index set, $\pi$ policies, $\beta$ the discount factor, and $\lambda$ the Lagrangian dual variable.
	
	\section{System Model and Problem Formulation}
	\label{sec:system_model}
	
	\subsection{5G NR Uplink System Model}
	\label{subsec:UL_model}
	We consider a 5G NR system in which a gNB equipped with $N_\text{ant}$ receive antennas serves $K$ User Equipments (UEs), each with $N_\text{UE}$ transmit antennas and $N_l \le \min(N_\text{ant}, N_\text{UE})$ spatial layers. The system operates with subcarrier spacing $\Delta f$ (numerology $\mu$), yielding a slot duration $T_\text{slot}$. The available bandwidth spans $N_\text{PRB}$ Physical Resource Blocks (PRBs), each comprising 12 contiguous subcarriers. We focus on \textit{uplink} link adaptation: the gNB selects the MCS for each UE's Physical Uplink Shared Channel (PUSCH) transmission, assuming a CP-OFDM waveform (transform precoding disabled) to enable multi-layer MIMO, and conveys it via the uplink grant (DCI Format~0\_0 or~0\_1)~\cite{3GPP_TS38212}, following the standard procedure of 3GPP TS~38.214 Section~6.1.4~\cite{3GPP_TS38214}.
	
	The gNB acquires channel state information (CSI) by processing PUSCH transmissions. Let $\mathbf{H}_{t,k,n} \in \mathbb{C}^{N_\text{ant} \times N_l}$ denote the uplink channel matrix for UE $k$ on subcarrier $n$ at slot $t$. The received uplink signal vector at the gNB is:
	\begin{equation}\label{eq:received_signal}
		\mathbf{y}_{t,k,n} = \mathbf{H}_{t,k,n} \mathbf{x}_{t,k,n} + \mathbf{n}_{t,k,n},
	\end{equation}
	where $\mathbf{x}_{t,k,n} \in \mathbb{C}^{N_l}$ is the transmitted signal vector across $N_l$ spatial layers and $\mathbf{n}_{t,k,n} \sim \mathcal{CN}(\mathbf{0}, \sigma^2\mathbf{I})$ is additive white Gaussian noise. From~(\ref{eq:received_signal}), the gNB estimates the channel using uplink Demodulation Reference Signals (DMRS)~\cite{3GPP_TS38211} via linear estimation (e.g., MMSE), obtaining $\hat{\mathbf{H}}_{t,k,n}$ together with a wideband SINR estimate $\gamma_{t,k}$ and Reference Signal Received Power (RSRP) for each scheduled UE.
	
	In each slot $t$, the MAC scheduler allocates $N_{\text{PRB},k}$ PRBs to each scheduled UE and the link adaptation mechanism selects an MCS index $m_{t,k} \in \mathcal{M} = \{0, 1, \ldots, 27\}$ ($|\mathcal{M}| = 28$ entries), determining the modulation order $Q_m$ and target code rate $r$ per the 64QAM MCS table~\cite[Table~5.1.3.1-1]{3GPP_TS38214}. The selected MCS is signaled to the UE via the 5-bit MCS field in the uplink grant~\cite{3GPP_TS38212}; the UE encodes its PUSCH transport block accordingly. The Transport Block Size (TBS) $B(m, N_{\text{PRB},k})$ is computed as a function of MCS index and allocated PRBs per the procedure in~\cite{3GPP_TS38214}. After PUSCH reception, the gNB decodes the transport block and verifies its CRC, obtaining an immediate transmission outcome without the signaling delay of downlink HARQ-ACK reporting via PUCCH:
	\begin{equation}\label{eq:harq_outcome}
		o_{t,k} = \begin{cases} 1 \;\text{(ACK)}, & \text{if CRC passes at the gNB} \\ 0 \;\text{(NACK)}, & \text{otherwise.} \end{cases}
	\end{equation}
	The instantaneous per-UE uplink throughput is $R_{t,k} = B(m_{t,k}, N_{\text{PRB},k}) \cdot o_{t,k} \,/\, T_\text{slot}$, where $o_{t,k}$ is the HARQ outcome~(\ref{eq:harq_outcome}). This uplink formulation offers a key advantage over downlink link adaptation: the gNB both selects the MCS and directly observes the decoding outcome, closing the feedback loop within a single entity without relying on TDD channel reciprocity or UE-reported CQI.
	
	\subsection{Conventional OLLA}
	\label{subsec:olla}
	Commercial 5G systems employ a two-loop link adaptation structure:
	\begin{enumerate}
		\item \textit{Inner loop (ILLA):} Maps the wideband SINR estimate $\gamma_{t,k}$ (measured by the gNB receiver) to an MCS index via a static Lookup Table (LUT):
		\begin{equation}\label{eq:illa}
			m_{t,k} = \text{LUT}\!\left(\min(\gamma_{t,k}, \gamma_\text{max}) + \Delta_{t,k}\right),
		\end{equation}
		where $\gamma_\text{max} = 25.99$\,dB is the SINR saturation threshold (the highest entry in the MCS lookup table) that prevents over-aggressive MCS selection at high SNR, and the lowest entry is $\gamma_\text{LUT}^{\min} = -4.57$\,dB.
		
		\item \textit{Outer loop (OLLA):} Maintains a per-UE SINR offset $\Delta_{t,k}$ that is adjusted based on HARQ feedback:
		\begin{equation}\label{eq:olla_update}
			\Delta_{t+1,k} = \Delta_{t,k} + \begin{cases} \delta_\text{up}, & \text{if } o_{t,k} = 1 \;\text{(ACK)} \\ -\delta_\text{down}, & \text{if } o_{t,k} = 0 \;\text{(NACK)} \end{cases}
		\end{equation}
		Under stationarity, the expected drift vanishes, yielding the well-known equilibrium BLER originally derived for outer-loop SIR control in~\cite{Sampath1997} and adopted by modern OLLA~\cite{BlanquezCasado2016eOLLAAE}:
		\begin{equation}\label{eq:olla_eq_bler}
			\varepsilon_\text{eq} = \frac{\delta_\text{up}}{\delta_\text{up} + \delta_\text{down}}.
		\end{equation}
	\end{enumerate}
	The offset is initialized at $\Delta_{0,k} = 0$\,dB. For the default step sizes $\delta_\text{up} = 0.1$\,dB and $\delta_\text{down} = 1.0$\,dB, (\ref{eq:olla_eq_bler}) yields $\varepsilon_\text{eq} \approx 9.1\%$.
	
	While robust and widely deployed, OLLA has two fundamental limitations. First, the update~(\ref{eq:olla_update}) is a first-order integral controller with scalar state $\Delta_t$ driven by a single bit of feedback per slot (ACK or NACK), ignoring the rich uplink channel state available at the gNB, including per-subcarrier channel estimates from PUSCH DMRS processing, temporal fading statistics, and received signal power. Second, the convergence rate is governed by the step sizes: smaller steps ensure stability but slow adaptation, while larger steps enable faster tracking at the cost of offset oscillations~\cite{BlanquezCasado2016eOLLAAE,Mota:GW:2019}. This trade-off is particularly acute in high-Doppler scenarios where the channel coherence time approaches the OLLA convergence time.
	
	\subsection{Constrained Optimization Formulation}
	\label{subsec:constrained_op}
	The link adaptation objective is to maximize expected discounted throughput subject to a long-term BLER constraint:
	\begin{align}
		\max_{\pi} \quad & (1-\beta)\,\mathbb{E}_\pi\!\left[ \sum_{t=0}^{\infty} \beta^t \, R_{t} \right] \label{eq:cmdp_obj} \\
		\text{s.t.} \quad & (1-\beta)\,\mathbb{E}_\pi\!\left[ \sum_{t=0}^{\infty} \beta^t\, c_t \right] \le \varepsilon_\text{target}, \label{eq:cmdp_bler}
	\end{align}
	where $\pi$ is the control policy mapping observations to MCS selections, $\beta \in (0,1)$ is the discount factor, $R_t$ is the per-slot throughput reward (instantiated in Section~\ref{subsec:state_obs}), $o_t \in \{0,1\}$ denotes the binary HARQ feedback (ACK/NACK, distinct from the observation vector $\mathbf{o}_t$), $c_t = \mathbb{I}(o_t = 0)$ is the corresponding constraint cost, and $\varepsilon_\text{target}$ is the maximum allowable BLER. The $(1-\beta)$ normalization follows the standard CMDP convention~\cite[\S2.2]{Altman1999CMDP}: under ergodicity, $(1-\beta)\mathbb{E}[\sum \beta^t f_t] \rightarrow \mathbb{E}_\text{ss}[f_t]$ as $\beta \rightarrow 1$ for any bounded per-slot quantity $f_t$~\cite[Remark~2.1]{Altman1999CMDP}, so the objective represents the average per-slot throughput and the constraint directly bounds the steady-state BLER. This is a Constrained Markov Decision Process (CMDP)~\cite{Altman1999CMDP}.
	
	Although link adaptation is a continuing task, the discounted objective is a standard and well-justified choice: the discount factor induces an effective planning horizon of $1/(1-\beta)$ steps matched to the timescale of channel dynamics, and the discounted-optimal policy converges to the average-optimal policy as $\beta \rightarrow 1$ under ergodicity~\cite{Altman1999CMDP}. In practice, the gNB observes only noisy channel estimates, quantized signal quality indicators, and binary HARQ feedback, rather than the true state. Section~\ref{sec:learned_olla} addresses this partial observability by extending the CMDP to a POMDP framework with history-enriched observations and Lagrangian constraint enforcement.

	\section{LOLLA Framework}
	\label{sec:learned_olla}
	As noted in Section~\ref{subsec:constrained_op}, the true system state, comprising the instantaneous channel realization, inter-cell interference, and UE buffer occupancy, is not directly accessible to the gNB. We therefore model the constrained link adaptation problem as a Partially Observable Markov Decision Process (POMDP)~\cite{Kaelbling1998PlanningAA}, defined by the tuple $\langle S, A, T, R, \Omega, O, \beta \rangle$, where $S$ is the latent state space (channel, interference, and buffer conditions), $A$ the continuous SINR offset action space, $\Omega$ the PHY/MAC observation space accessible to the gNB, $T(s, a, s') = \Pr(s' \mid s, a)$ the state-transition probability, $O(s', a, o) = \Pr(o \mid s', a)$ the observation-emission probability, $R$ the per-slot throughput reward, and $\beta \in (0,1)$ the discount factor. Each element is instantiated below.
	
	Because the agent cannot condition on $s \in S$ directly, an optimal POMDP policy must integrate information over the observation history $\mathbf{o}_{1:t}$ to maintain a sufficient statistic, the \textit{belief state} $b_t(s) = \Pr(s_t{=}s \mid \mathbf{o}_{1:t},\, \mathbf{a}_{0:t-1})$, over the latent dynamics~\cite{Kaelbling1998PlanningAA}. We approximate this belief state through a fixed-length observation window that embeds recent MCS indices, ACK/NACK outcomes, and a running BLER estimate directly into the observation vector (Section~\ref{subsec:state_obs}), enabling a reactive feedforward policy to capture the most salient temporal patterns without recurrent state. The BLER constraint~(\ref{eq:cmdp_bler}) is enforced via Lagrangian relaxation (Section~\ref{subsec:lagrangian_bler}).
	
	\subsection{POMDP Formulation}
	\label{subsec:state_obs}
	For notational clarity, we drop the UE index $k$ in this section; the formulation applies independently to each scheduled UE, and the multi-UE batched extension is described in Section~\ref{subsec:multi-UE-batched-control}.
	
	\textbf{Observation space.} The agent observes $\mathbf{o}_t \in \mathbb{R}^{d_o}$ with $d_o = 13$, constructed from slot-level PHY/MAC telemetry (the real-time data path is detailed in Section~\ref{sec:system_arch}):
	\begin{equation}\label{eq:obs}
		\mathbf{o}_t = [\mathbf{f}_\text{ch},\; \mathbf{f}_\text{sig},\; \mathbf{h}_\text{mcs},\; \mathbf{h}_\text{ack},\; \bar{\varepsilon}_t]^\top.
	\end{equation}
	\begin{itemize}
		\item \textbf{Channel features} $\mathbf{f}_\text{ch} \in \mathbb{R}^4$. The uplink channel estimate $\hat{\mathbf{H}} \in \mathbb{C}^{N_\text{ant} \times N_l \times N_\text{SC} \times N_\text{DMRS}}$ is compressed into four statistics: mean channel gain $\mu_{|H|}$ (path loss), standard deviation of per-subcarrier gains $\sigma_{|H|}$ (frequency selectivity), 10th-percentile gain $P_{10}(|H|)$ (deep-fade severity), and cross-slot Pearson correlation $\rho_t$ (instantaneous Doppler indicator). These capture per-slot channel state entirely discarded by OLLA's one-bit feedback.
		\item \textbf{Signal quality features} $\mathbf{f}_\text{sig} \in \mathbb{R}^2$. The wideband SINR $\gamma_t$ (dB), measured directly by the gNB L1 PHY pipeline, which is the primary LUT input, and RSRP (dBm), which disambiguates low SINR due to path loss from low SINR due to interference.
		\item \textbf{MCS history} $\mathbf{h}_\text{mcs} = [m_{t-1}, m_{t-2}, m_{t-3}]^\top / 27 \in [0,1]^3$. The three most recent MCS selections (default $0.5$), enabling detection of MCS oscillations, a key OLLA failure mode under rapid channel variation.
		\item \textbf{HARQ history} $\mathbf{h}_\text{ack} = [o_{t-1}, o_{t-2}, o_{t-3}]^\top \in \{0,1\}^3$. The corresponding three-step binary outcomes; together with $\mathbf{h}_\text{mcs}$, these form action-outcome pairs $(m_{t-i}, o_{t-i})$ that subsume OLLA's single-bit feedback.
		\item \textbf{Running BLER} $\bar{\varepsilon}_t \in [0,1]$. A sliding-window average (window size $W = 100$ slots) that gives the agent direct observability of the constraint state relative to $\varepsilon_\text{target}$, accelerating Lagrangian convergence.
	\end{itemize}
	
	\textbf{Action space.} The action $a_t = \delta_t \in [-\delta_\text{max}, +\delta_\text{max}]$ is a continuous scalar SINR offset (in dB). The final MCS is selected via the same standards-compliant pipeline~(\ref{eq:illa}) used by OLLA:
	\begin{equation}\label{eq:lolla_action}
		m_t = \text{LUT}\!\left( \min(\gamma_t, \gamma_\text{max}) + \delta_t \right),
	\end{equation}
	where $\gamma_t$ is the wideband SINR (in dB) and $\gamma_\text{max}$ is the SINR cap. Fig.~\ref{fig:fig2_method_comparison} contrasts the three action formulations: (a)~conventional OLLA adjusts its offset $\Delta_t$ via a scalar ACK/NACK staircase, consuming only one bit of feedback per slot; (b)~discrete RL bypasses the SINR-to-MCS LUT entirely and selects $m_t \in \{0,\ldots,27\}$ directly from the policy, forgoing the structured SINR-to-MCS mapping as a prior; (c)~LOLLA retains the identical SINR-capped LUT pipeline while replacing the staircase with a policy-generated offset $\delta_t$ conditioned on the full 13-dimensional observation. This residual formulation~\cite{Johannink:ICRA:2019} ensures that $\delta_t \approx 0$ recovers the SINR-only baseline (inner-loop LUT without outer-loop adaptation), providing a strong inductive bias for safe operation and faster convergence compared to direct MCS prediction. Crucially, this shared pipeline establishes a formal performance guarantee:
	
	\putFig{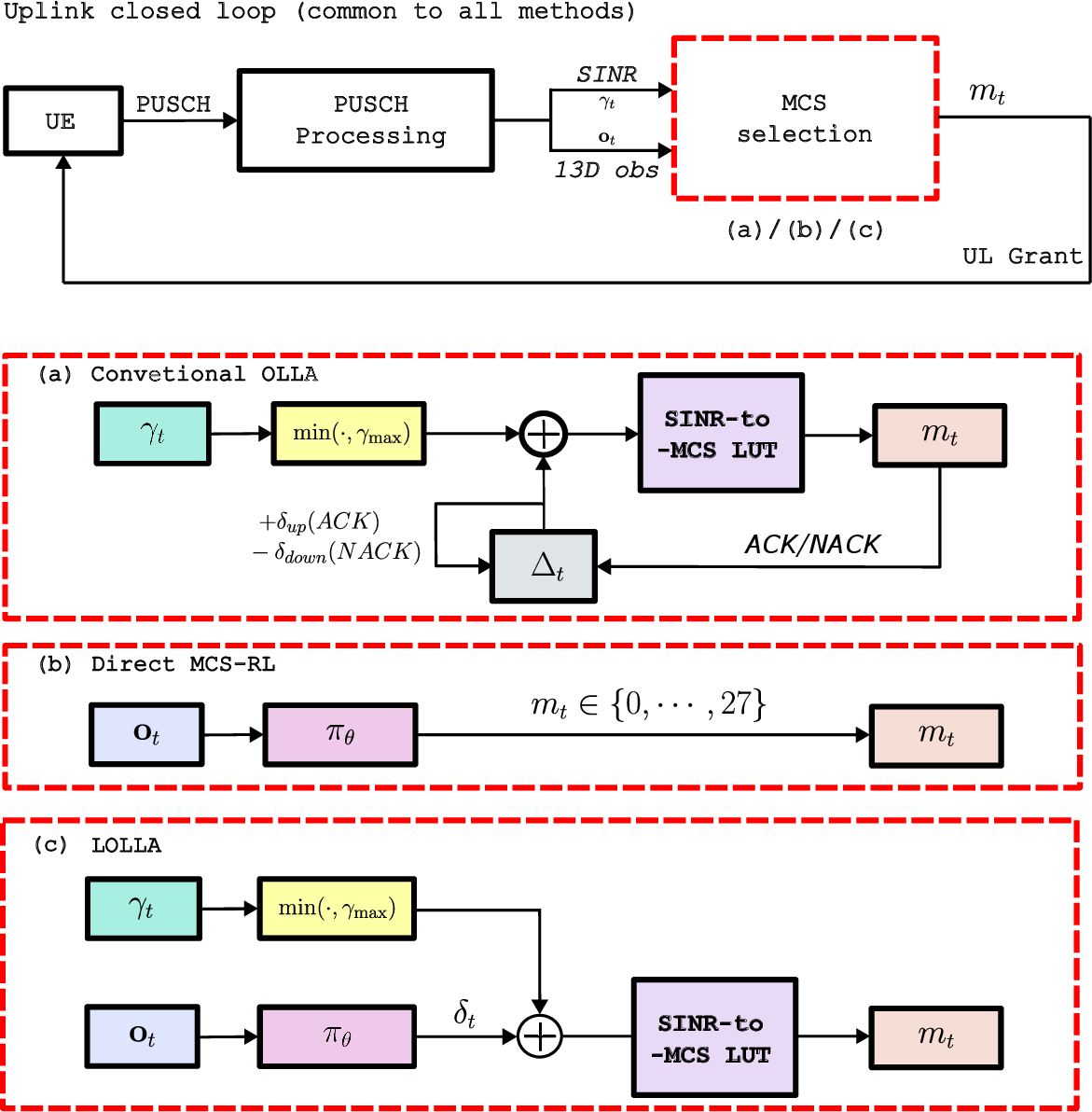}{Link adaptation action formulations. (a)~OLLA: ACK/NACK staircase adjusts $\Delta_t$. (b)~Direct MCS-RL: categorical policy selects MCS index directly, bypassing the LUT. (c)~LOLLA (proposed): continuous offset $\delta_t$ modulates the standards-compliant LUT, conditioned on PHY/MAC observations~$\mathbf{o}_t$.}{3.45}
	
	\begin{proposition}[Performance Dominance]\label{prop:dominance}
		Let $\Pi_\text{OLLA}$ denote the class of OLLA policies parameterized by $(\delta_\text{up}, \delta_\text{down})$, and let $\Pi_\text{RL}$ denote the class of all measurable policies mapping observation histories $\mathbf{o}_{1:t}$ to offsets $\delta_t \in [-\delta_\text{max}, +\delta_\text{max}]$. Assume the wideband SINR satisfies $\gamma_t \ge \gamma_\text{min}$ a.s.\ for a receiver sensitivity $\gamma_\text{min} \le \gamma_\text{LUT}^{\min}$, where $\gamma_\text{LUT}^{\min}$ is the lowest SINR threshold in the MCS lookup table, and that $\delta_\text{max} \ge \gamma_\text{max} - \gamma_\text{min}$. Then every OLLA policy is realizable within $\Pi_\text{RL}$ (i.e., induces an identical MCS trajectory), and consequently
		\[
		\sup_{\pi \in \Pi_\text{RL}}\; (1-\beta)\,\mathbb{E}_\pi\!\left[\sum_{t=0}^{\infty} \beta^t \bar{r}_t\right] \;\ge\; \sup_{\pi \in \Pi_\text{OLLA}}\; (1-\beta)\,\mathbb{E}_\pi\!\left[\sum_{t=0}^{\infty} \beta^t \bar{r}_t\right].
		\]
	\end{proposition}
	
	The proof (Appendix~\ref{app:proof_dominance}) constructs an explicit embedding of any OLLA update rule within $\Pi_\text{RL}$. In practice, the feedforward MLP conditions on the current observation rather than the full history, incurring an approximation gap of $O(\varepsilon/(1-\beta))$ (Proposition~\ref{prop:duality}(iii)), where $\varepsilon$ jointly absorbs the policy parameterization error and the history-truncation error of the feedforward MLP class; however, the enriched 13-dimensional observation enables the learned policy to outperform OLLA through proactive rather than reactive adaptation.
	
	\textbf{Reward function.} We define the base throughput reward as:
	\begin{equation}\label{eq:base_reward}
		\bar{r}_t = \frac{R_t}{R_\text{norm}},
	\end{equation}
	where $R_t = B(m_t, N_\text{PRB}) \cdot o_t / T_\text{slot}$ is the per-slot effective throughput defined in Section~\ref{subsec:UL_model} and $R_\text{norm}>0$ is a fixed normalization constant chosen per configuration. Since $R_\text{norm}$ is a policy-independent positive constant, replacing $R_t$ by $\bar r_t$ in~(\ref{eq:cmdp_obj}) only rescales the objective by a constant and leaves both the optimal policy and the constraint~(\ref{eq:cmdp_bler}) unchanged; the resulting policy is further invariant to this scale through PPO's advantage normalization. As derived in Section~\ref{subsec:lagrangian_bler}, the Lagrangian relaxation of the BLER constraint yields the per-step training reward:
	\begin{equation}\label{eq:training_reward}
		r_t = \bar{r}_t - \lambda_t \cdot c_t = \begin{cases}
			R_t / R_\text{norm}, & \text{if } o_t = 1 \text{ (ACK)} \\[6pt]
			-\lambda_t, & \text{if } o_t = 0 \text{ (NACK)}
		\end{cases}
	\end{equation}
	where $\lambda_t \ge 0$ is the Lagrangian dual variable and $c_t = \mathbb{I}(o_t = 0) = 1 - o_t$ is the binary constraint cost. The flat NACK penalty $-\lambda_t$ (independent of MCS) ensures consistency with the binary constraint cost used in the dual variable update, forming a valid Lagrangian~\cite{Achiam2019BenchmarkingSE}.
	
	\subsection{Lagrangian BLER Constraint}
	\label{subsec:lagrangian_bler}
	To enforce the BLER constraint~(\ref{eq:cmdp_bler}) without manual penalty tuning, we adopt the Lagrangian relaxation framework for CMDPs~\cite{Altman1999CMDP,Achiam2019BenchmarkingSE}. Denoting $J_r(\pi) \triangleq (1-\beta)\,\mathbb{E}_\pi[\sum_t \beta^t \bar{r}_t]$ with $\bar{r}_t$ from~(\ref{eq:base_reward}) and $J_c(\pi) \triangleq (1-\beta)\,\mathbb{E}_\pi[\sum_t \beta^t c_t]$ as the $(1-\beta)$-normalized reward and constraint cost~\cite[\S2.2]{Altman1999CMDP}, the constrained objective is relaxed into an unconstrained saddle-point problem:
	\begin{equation}\label{eq:lagrangian}
		\mathcal{L}(\pi, \lambda) = (1-\beta)\,\mathbb{E}_\pi\!\left[\sum_{t=0}^{\infty} \beta^t \bigl(\bar{r}_t - \lambda\, c_t\bigr)\right] + \lambda\,\varepsilon_\text{target},
	\end{equation}
	where $\lambda \ge 0$ is the Lagrangian dual variable. Equivalently, this can be written as $\mathcal{L} = J_r(\pi) - \lambda\bigl(J_c(\pi) - \varepsilon_\text{target}\bigr)$, the standard Lagrangian for the constraint $J_c(\pi) \le \varepsilon_\text{target}$. Since the $(1-\beta)$ prefactor and the additive term $\lambda\varepsilon_\text{target}$ are both constant with respect to $\pi$, the per-step training reward is $r_t = \bar{r}_t - \lambda\, c_t$ as in~(\ref{eq:training_reward}). The policy $\pi$ maximizes $\mathcal{L}$ via PPO, while $\lambda$ is updated via dual gradient ascent on the constraint violation:
	\begin{equation}\label{eq:dual_update}
		\lambda_{k+1} = \text{clip}\!\left(\lambda_k + \eta \cdot (\bar{\varepsilon}_k - \varepsilon_\text{target}),\; 0,\; \lambda_\text{max}\right),
	\end{equation}
	where $k$ indexes PPO iterations (cf.\ Algorithm~\ref{alg:training}), $\eta$ is the dual learning rate, and $\bar{\varepsilon}_k \triangleq N_\text{NACK}/B$ is the rollout-batch BLER over the $B = TN$ transitions collected in iteration $k$, distinct from the per-slot running-window feature $\bar\varepsilon_t$ of Section~\ref{subsec:state_obs}. The clip to $[0, \lambda_\text{max}]$ ensures dual feasibility and numerical stability.
	
	Although $\delta_t$ is continuous, the SINR-to-MCS LUT (Section~\ref{subsec:state_obs}) maps the offset to one of $|\mathcal{M}| = 28$ MCS indices, so the effective action space is finite. The correctness of this approach rests on the following duality result.
	
	\begin{proposition}[Zero Duality Gap]\label{prop:duality}
		Consider a finite CMDP with $|S| < \infty$ states and $|\mathcal{M}| < \infty$ actions that approximates the link adaptation problem~(\ref{eq:cmdp_obj})--(\ref{eq:cmdp_bler}) via state-space discretization. Denote the discounted constraint cost $J_c(\pi) \triangleq (1-\beta)\,\mathbb{E}_\pi[\sum_t \beta^t c_t]$. Then:
		\begin{enumerate}
			\item[(i)] Strong duality holds for the Lagrangian~(\ref{eq:lagrangian}): $\max_\pi \min_{\lambda \ge 0} \mathcal{L}(\pi, \lambda) = \min_{\lambda \ge 0} \max_\pi \mathcal{L}(\pi, \lambda)$.
			\item[(ii)] There exists a saddle point $(\pi^*, \lambda^*)$ satisfying complementary slackness: $\lambda^* (J_c(\pi^*) - \varepsilon_\text{target}) = 0$.
			\item[(iii)] If additionally Slater's condition holds (i.e., $\exists\, \pi_0$ with $J_c(\pi_0) < \varepsilon_\text{target}$) and policies are parameterized by a function class $\Pi_\Theta$ that is $\varepsilon$-universal~\cite[Definition~1]{Paternain:NIPS:2019} (i.e., for every policy $\pi$ there exists $\theta$ with $\max_{s\in\mathcal{S}}\int_{\mathcal{A}}|\pi(a|s) - \pi_\theta(a|s)|\,da \le \varepsilon$), then the parametric duality gap, bounding both value sub-optimality and constraint violation, is $O(\varepsilon/(1-\beta))$.
		\end{enumerate}
	\end{proposition}
	
	The proof is in Appendix~\ref{app:proof_duality}: (i) follows from the minimax theorem on the convex-compact set of occupation measures~\cite[Theorems~3.2, 3.6]{Altman1999CMDP}; (ii) from Kuhn--Tucker conditions~\cite[Theorem~3.6(iii)]{Altman1999CMDP}; (iii) from~\cite{Paternain:NIPS:2019} under Slater's condition. Practically, the dual update~(\ref{eq:dual_update}) acts as a discrete-time integral controller on the constraint error $\bar{\varepsilon}_t - \varepsilon_\text{target}$~\cite{Stooke2020ResponsiveSI}, driving the aggregate BLER to the target without manual penalty calibration.
	
	This primal-dual formulation offers three additional properties compared to fixed penalty coefficients.
	
	\textbf{Target generality.} The Lagrangian decouples policy optimization from constraint enforcement: changing $\varepsilon_\text{target}$ modifies only the dual update without altering the policy architecture or training procedure. A single framework therefore supports requirements ranging from eMBB ($\varepsilon_\text{target} \approx 10\%$) to URLLC ($\varepsilon_\text{target} < 1\%$) by specifying the desired target before training.
	
	\textbf{Asymmetric enforcement dynamics.} From~(\ref{eq:dual_update}), the dual gradient $g_t = \bar{\varepsilon}_t - \varepsilon_\text{target}$ is bounded in $[-\varepsilon_\text{target},\, 1 - \varepsilon_\text{target}]$, so $\lambda$ increases up to $(1 - \varepsilon_\text{target})/\varepsilon_\text{target}$ times faster upon constraint violation than it decreases during satisfaction (e.g., ${\approx}10\times$ at $\varepsilon_\text{target} = 9.1\%$), a desirable ``fast up, slow down'' property for safety-critical link adaptation~\cite{Stooke2020ResponsiveSI}. This mirrors OLLA's asymmetric step sizes ($\delta_\text{down}/\delta_\text{up} = 10$)~\cite{BlanquezCasado2016eOLLAAE}, but emerges automatically from the constraint structure without manual tuning.
	
	\textbf{Non-uniform per-SNR BLER allocation.} Because the ACK reward $R_t/R_\text{norm}$ in~(\ref{eq:training_reward}) grows with MCS index while the flat NACK penalty $-\lambda$ is fixed, the penalty-to-reward ratio decreases at high SNR, incentivizing the agent to tolerate higher BLER where aggressive MCS selection yields the largest throughput gain, and to enforce lower BLER at low SNR where the penalty dominates. A throughput-proportional penalty (scaling $-\lambda$ with the same factor) would instead produce uniform BLER allocation analogous to conventional OLLA. The aggregate BLER converges to $\varepsilon_\text{target}$ through dual feedback on $\lambda$, while the per-SNR allocation concentrates the error budget where gains are largest. Under a continuous MCS relaxation, assuming each per-SNR throughput function $T_i$ is differentiable and concave in $\varepsilon_i$, this is throughput-optimal by the KKT equal-marginal condition $T_i'(\varepsilon_i^*) = \mu$~\cite[\S5.5.3]{Boyd2004}.
	
	\subsection{PPO Policy Architecture}
	\label{subsec:ppo}
	We train a feedforward actor-critic network with Proximal Policy Optimization (PPO)~\cite{Schulman2017ProximalPO}. Although the underlying system is partially observable, the observation vector $\mathbf{o}_t$ already embeds a fixed-length history window (three-step MCS and ACK histories, running BLER) that captures the most salient temporal patterns. A feedforward (MLP) policy conditioned on this enriched observation achieves strong performance without the additional complexity of recurrent state management.
	
	\textbf{Network architecture.} The actor and critic are two independent two-layer MLPs with $\tanh$ activations and hidden dimension $n_h$. The actor outputs the mean of a Gaussian policy $\pi_\theta(\delta_t \mid \mathbf{o}_t) = \mathcal{N}(\mu_\theta(\mathbf{o}_t),\, \sigma^2)$, where $\ln \sigma$ is a state-independent learnable parameter; the critic shares the same architecture with independent parameters, producing a scalar value estimate $\hat{V}_\theta(\mathbf{o}_t)$. Weights are orthogonally initialized~\cite{Saxe2013ExactST} with gain $0.01$ for the actor output, ensuring near-zero initial offsets $\delta_t \approx 0$ so that the initial policy approximates the SINR-only baseline.
	
	\textbf{PPO objective.} The policy is updated by maximizing the clipped surrogate objective~\cite{Schulman2017ProximalPO}:
	\begin{equation}\label{eq:ppo_clip}
		L^\text{CLIP}(\theta) = \hat{\mathbb{E}}_t\!\left[\min\!\left(
		\rho_t\, \hat{A}_t,\;
		\text{clip}(\rho_t,\, 1{-}\epsilon,\, 1{+}\epsilon)\, \hat{A}_t
		\right)\right],
	\end{equation}
	where $\rho_t = \pi_\theta / \pi_{\theta_\text{old}}$ is the importance sampling ratio and $\hat{A}_t$ is the advantage estimated via GAE($\lambda_\text{GAE}$)~\cite{Schulman2015} with done-masked bootstrapping. The full loss combines the clipped surrogate~(\ref{eq:ppo_clip}), a clipped value function loss~\cite{Engstrom2020}, and a Gaussian entropy bonus $H[\pi_\theta] = \frac{1}{2}\ln(2\pi e\, \sigma^2)$:
	\begin{equation}\label{eq:ppo_loss}
		L(\theta) = -L^\text{CLIP}(\theta) + c_v \, L^\text{VF}(\theta) - c_e \, H[\pi_\theta].
	\end{equation}
	The entropy term in~(\ref{eq:ppo_loss}) prevents premature collapse of $\sigma$, which is critical for our one-dimensional action space where the policy can converge to a near-deterministic mapping without exploration pressure.
	
	\subsection{Training Procedure}
	\label{subsec:training_procedure}
	Algorithm~\ref{alg:training} presents the unified training loop. At each iteration $k$, the agent collects $B = TN$ transitions (rollout length $T$ across $N$ parallel environments), performs $K_\text{epoch}$ PPO epochs over $M$ mini-batches, and adjusts the dual variable $\lambda$ based on the observed BLER. The loop runs for $K_\text{total} = \lfloor T_\text{total} / B \rfloor$ iterations. All hyperparameters, including the Adam~\cite{Kingma2015} optimizer settings, are listed in Table~\ref{tab:sim_params}.
	
	\begin{algorithm}[t]
		\caption{LOLLA Training}
		\label{alg:training}
		\begin{algorithmic}[1]
			\Require Policy $\pi_\theta$, initial dual variable $\lambda_0$, hyperparameters (Table~\ref{tab:sim_params})
			\State Initialize running-statistics normalizer $\mathcal{R}$
			\For{$k = 0, 1, \ldots, K_{\text{total}} - 1$}
			\For{$t = 0, \ldots, T{-}1$ across $N$ environments} \Comment{Rollout}
			\State Update $\mathcal{R}$ with $\mathbf{o}_t$
			\State $\tilde{\mathbf{o}}_t \gets \text{clip}\bigl((\mathbf{o}_t {-} \hat{\boldsymbol{\mu}}) / \sqrt{\hat{\boldsymbol{\sigma}}^2 {+} \epsilon},\; {-}c_o,\; c_o\bigr)$
			\State $\delta_t,\, \log\pi_t,\, \hat{V}_t \gets \pi_\theta(\tilde{\mathbf{o}}_t)$
			\State $m_t \gets \Call{Lut}{\min(\gamma_t, \gamma_{\max}) + \delta_t}$
			\State Execute $m_t$; observe $r_t$, $\mathbf{o}_{t+1}$, $d_{t+1}$
			\EndFor
			\State $\{\hat{A}_t\} \gets \Call{GAE}{\lambda_{\text{GAE}}}$ with done-masked bootstrap \Comment{Sec.~\ref{subsec:ppo}}
			\State $\alpha_k \gets \max\bigl(f_{\text{min}},\; 1 {-} k/K_{\text{total}}\bigr) \cdot \alpha_0$ \Comment{LR with floor}
			\For{epoch $= 1, \ldots, K_{\text{epoch}}$} \Comment{PPO update}
			\State Flatten and randomly shuffle all $B = TN$ transitions
			\State Partition into $M$ mini-batches
			\For{each mini-batch $\mathcal{S}$}
			\State Recompute $\pi_\theta$, $\hat{V}_\theta$ on $\mathcal{S}$
			\State $\hat{A}_{\mathcal{S}} \gets \bigl(\hat{A}_{\mathcal{S}} {-} \operatorname{mean}(\hat{A}_{\mathcal{S}})\bigr) \big/ \bigl(\operatorname{std}(\hat{A}_{\mathcal{S}}) {+} \epsilon\bigr)$
			\State $\theta \gets \theta - \alpha_k \, \nabla_\theta L(\theta)$ after clipping $\|\nabla_\theta L\|_2$ to $g_{\max}$
			\EndFor
			\EndFor
			\State $\bar{\varepsilon}_k \gets N_{\text{NACK}} / B$ \Comment{Lagrangian dual update}
			\State $\lambda_{k+1} \gets \bigl[\lambda_k + \eta\,(\bar{\varepsilon}_k - \varepsilon_{\text{target}})\bigr]_0^{\lambda_{\max}}$ \Comment{Sec.~\ref{subsec:lagrangian_bler}}
			\State Set NACK penalty $\gets \lambda_{k+1}$
			\EndFor
		\end{algorithmic}
	\end{algorithm}
	
	Three aspects of the training setup merit further discussion.
	
	\textbf{Episode structure and domain randomization} (line~3). Link adaptation is a continuing task with no natural termination; episodes are truncated at a fixed horizon of $T_\text{ep}$ steps. To promote generalization, we apply SNR domain randomization~\cite{Tobin2017DomainRF}: the operating SNR is drawn uniformly from a prescribed range at episode start and re-sampled with probability $p_\text{SNR}$ per slot, exposing the agent to diverse link conditions within each episode. The channel model and Doppler frequency are fixed per training run; cross-scenario generalization is evaluated in Section~\ref{subsec:generalization}.
	
	\textbf{Observation normalization} (line~5). Each feature is normalized online using a Welford running mean/variance estimator~\cite{Welford1962NoteOA} ($\epsilon = 10^{-8}$), with statistics updated at every environment step. Advantages are additionally zero-mean/unit-variance normalized within each mini-batch (line~16), a standard PPO practice that reduces sensitivity to reward scale~\cite{Engstrom2020}.
	
	\textbf{Learning rate floor} (line~10). The learning rate is linearly annealed from $\alpha_0$, but a minimum fraction $f_\text{min} \in [0, 1]$ ensures it never falls below $f_\text{min} \cdot \alpha_0$. Without this floor, $\lambda$ may still be converging in late training while the near-zero learning rate freezes the policy, preventing the BLER constraint from being satisfied.
	
	\begin{remark}[Computational Complexity]\label{rem:complexity}
		\textit{Training.} Each of the $T_\text{total}/B$ iterations collects $B = TN$ transitions and performs $K_\text{epoch}$ sweep epochs, each partitioned into $M$ minibatch gradient steps of size $B/M$, yielding a total cost of $O(K_\text{epoch} \cdot T_\text{total} \cdot |\Theta|)$, where $|\Theta| = O(d_o n_h + n_h^2)$ is the parameter count of the actor-critic network. The Lagrangian dual update and GAE computation are both $O(B)$, dominated by the network passes.
		\textit{Inference.} Per-slot inference requires one actor forward pass of cost $O(d_o n_h + n_h^2)$ plus a constant-time LUT lookup, and scales linearly with the number of scheduled UEs $K$ under batched execution.
	\end{remark}
	
	\section{Real-Time System Architecture}
	\label{sec:system_arch}
	This section describes how LOLLA (Section~\ref{sec:learned_olla}) is realized as a closed-loop dApp on a GPU-accelerated 5G NR stack. The design builds on the E3 dApp framework of Villa~\emph{et al.}~\cite{Villa2025ProgrammableAG}, which established the first GPU-native platform for slot-level PHY/MAC telemetry and specifies an E3 control path for actuation. Its realized dApps, however, are open-loop, and the reference explicitly notes that \textit{control logic is not implemented}~\cite{Villa2025ProgrammableAG}. LOLLA realizes the closed control loop this framework leaves unimplemented: a dual-agent E3 architecture bridges DU-Low and DU-High through zero-copy shared memory, and a four-stage inference backend progression culminating in a Triton C-API binding meets the per-slot latency budget. Fig.~\ref{fig:fig_architecture} overviews the resulting closed-loop dApp; the rest of this section details its components along the per-slot data path.
	
	\putFig{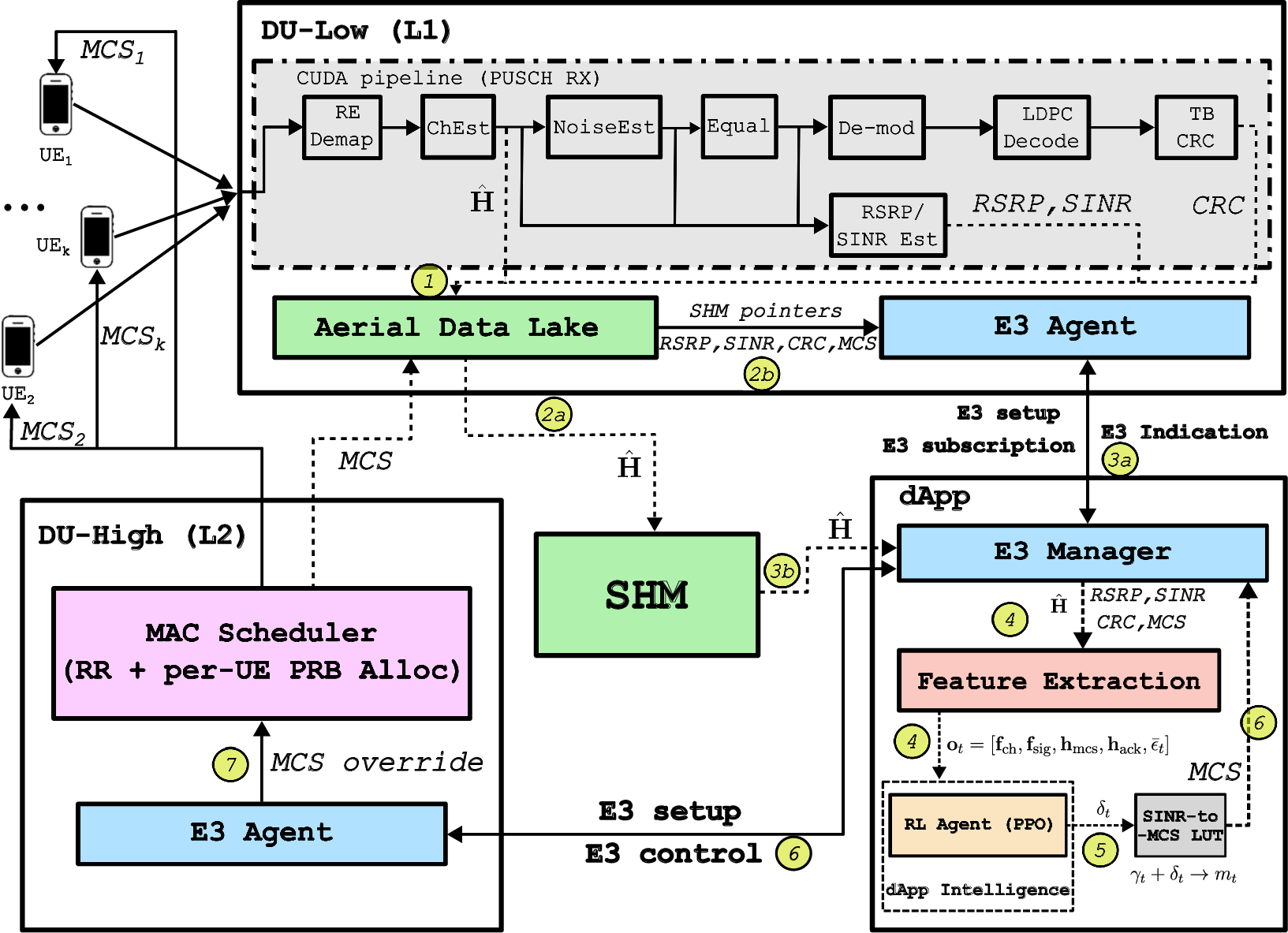}{LOLLA system architecture. The L1 E3 Agent exports PHY telemetry via the Aerial Data Lake (shared memory) and ZMQ; the dApp performs feature extraction and policy inference; the L2 E3 Agent delivers per-UE MCS overrides to the MAC scheduler.}{3.45}
	
	\subsection{Dual-Agent Closed-Loop Architecture}
	\label{subsec:closed-loop-dApp}
	The E3 framework~\cite{Villa2025ProgrammableAG} exposes PHY/MAC telemetry to in-gNB dApps through \textit{E3 Agents} coordinated by a co-located \textit{E3 Manager} under the E3 Application Protocol (E3AP)~\cite{LACAVA2025111342}. Bulk PHY tensors traverse the Aerial Data Lake (ADL), a zero-copy POSIX shared-memory region with a double-buffered ping-pong layout, while per-slot scalars and synchronization fields ride a lightweight ZeroMQ (ZMQ) transport carrying E3 Indication and E3 Control messages.
	
	LOLLA operates a \emph{dual-agent} configuration (Fig.~\ref{fig:fig_architecture}) that closes the control loop:
	\begin{itemize}
		\item \textit{L1 E3 Agent (DU-Low).} Integrated into the CUDA PHY pipeline, it stages each per-slot channel estimate $\hat{\mathbf{H}}$ into the active ADL ping-pong buffer and publishes the wideband SINR $\gamma_t$ (Section~\ref{subsec:UL_model}), RSRP, per-TB CRC outcomes, timing identifiers (SFN, slot, RNTI), and the SHM pointers as an E3 Indication on a ZMQ \textsc{pub} socket.
		\item \textit{L2 E3 Agent (DU-High).} A new agent co-located with the MAC scheduler that exposes a ZMQ \textsc{rep} endpoint accepting E3 Control messages with single-UE or batched multi-UE MCS overrides; a registered callback applies them to the next uplink grant, providing the control-plane actuation path.
	\end{itemize}
	On the dApp side, the E3 Manager opens a ZMQ \textsc{dealer} socket connected to the L2 Agent's \textsc{rep} endpoint, dispatching each E3 Control as a single frame without waiting for a reply; this fire-and-forget path achieves a measured send latency of $\sim 23$--$25\,\mu$s (Section~\ref{subsec:latency_breakdown}, Op~9) while the L1 \textsc{pub}-\textsc{sub} telemetry path runs on independent sockets.
	
	At each slot~$t$, the closed loop proceeds as:
	\begin{enumerate}
		\item The CUDA PHY pipeline executes the uplink receive chain (channel estimation, noise estimation, MMSE equalization with soft demapping, LDPC decoding, CRC check) on the received PUSCH, producing the telemetry tuple $(\hat{\mathbf{H}}, \gamma_t, \mathrm{RSRP}, \mathrm{CRC})$.
		\item $\hat{\mathbf{H}}$ is copied into the active ADL ping-pong shared-memory buffer~(2a); upon completion, the ADL notifies the L1 Agent with the SHM pointers (the active ping-pong buffer index, the in-buffer write offset, and the tensor shape) together with the scalar fields ($\gamma_t$, RSRP, per-TB CRC, and timing identifiers)~(2b).
		\item The L1 Agent publishes the corresponding E3 Indication, carrying those SHM pointers and scalar fields, over its ZMQ \textsc{pub} socket, which the E3 Manager receives~(3a); the E3 Manager then memory-maps $\hat{\mathbf{H}}$ from the indicated shared-memory segment using those pointers, exposing the tensor in its address space without any data copy~(3b).
		\item The dApp extracts the four channel statistics $\mathbf{f}_\text{ch}$ from $\hat{\mathbf{H}}$ on the CPU (Section~\ref{subsec:shared-memory}) and assembles $\mathbf{o}_t$ together with the indication scalars and the per-UE state buffers.
		\item The inference backend (Section~\ref{subsec:inference-pipeline}) runs the actor network on $\mathbf{o}_t$ and emits the SINR offset $\delta_t$.
		\item The dApp maps $\delta_t$ through the SINR-to-MCS LUT~(\ref{eq:lolla_action}) and dispatches the resulting per-UE MCS override to the L2 Agent over the ZMQ \textsc{dealer} socket.
		\item The L2 Agent's registered callback applies the override on the next uplink grant; the subsequent PUSCH outcome feeds back into Step~1 as fresh telemetry, yielding $\mathbf{o}_{t+1}$ and closing the loop.
	\end{enumerate}
	
	Steps~6 and~7 close the loop beyond the open-loop inference dApps of~\cite{Villa2025ProgrammableAG}: the dApp now actuates control and observes its consequences. Because the dApp interacts with the gNB only through E3AP messages~\cite{LACAVA2025111342} sent to configurable endpoint addresses, the same binary runs against either a GPU-accelerated PHY simulator or a production gNB by reconfiguring those addresses.
	
	\subsection{Shared-Memory Telemetry and Feature Extraction}
	\label{subsec:shared-memory}
	The uplink channel estimate $\hat{\mathbf{H}} \in \mathbb{C}^{N_\text{ant} \times N_l \times N_\text{SC} \times N_\text{DMRS}}$ is staged into POSIX shared memory by the ADL following Ops~1--3 of~\cite[Table~1]{Villa2025ProgrammableAG}; for the $4 \times 2$ MIMO configuration of Table~\ref{tab:sim_params} ($N_\text{SC} = 3276$, $N_\text{DMRS} = 1$, complex64), $\hat{\mathbf{H}}$ occupies approximately $205$\,KiB per UE per slot in single-UE mode. Rather than uploading the raw tensor back to the inference GPU, the dApp compresses it in place on the CPU into the four channel statistics $\mathbf{f}_\text{ch} = [\mu_{|H|}, \sigma_{|H|}, P_{10}(|H|), \rho_t]$ defined in Section~\ref{subsec:state_obs} via a vectorized C++ module that parallelizes across UEs with OpenMP; the cross-slot correlation $\rho_t$ uses an RNTI-keyed cache of the previous slot's per-subcarrier magnitude profile.
	
	The 13-dimensional observation $\mathbf{o}_t$~(\ref{eq:obs}) is then assembled from $\mathbf{f}_\text{ch}$, the signal-quality features $\mathbf{f}_\text{sig} = [\gamma_t, \mathrm{RSRP}]^\top$ carried as scalars in the same E3 Indication, and the per-UE history buffers $(\mathbf{h}_\text{mcs}, \mathbf{h}_\text{ack}, \bar{\varepsilon}_t)$, which are updated \emph{before} observation construction so that $\mathbf{o}_t$ already reflects the previous slot's outcome. The resulting $52$-byte vector represents an ${\sim}4{,}000\times$ compression of $\hat{\mathbf{H}}$ that removes bulk-PHY-tensor uploads from the per-slot critical path. Because the GPU-accelerated PHY simulator and the production ARC-OTA gNB execute the same CUDA baseband kernels~\cite{Villa2025ProgrammableAG}, with only the RF front-end replaced by calibrated 3GPP TDL channel models (Section~\ref{subsec:setup}), the PHY-derived components of $\mathbf{o}_t$ are produced by identical kernels at training and at deployment, while $(\mathbf{h}_\text{mcs}, \mathbf{h}_\text{ack}, \bar{\varepsilon}_t)$ are dApp-internal and PHY-agnostic.
	
	\subsection{Model Export and Inference Pipeline}
	\label{subsec:inference-pipeline}
	
	\textbf{Model export.} The trained actor is exported to ONNX with optional FP16 TensorRT compilation; the observation running-mean and running-variance from training (Section~\ref{subsec:training_procedure}) are baked into the graph as constant tensors, so inference applies the same normalization as training. Dynamic batch axes let one exported model serve both single-UE and multi-UE inference, and a Triton model repository is generated automatically.
	
	\textbf{Inference backend progression.} Inference lies on the per-slot critical path, where the gRPC round-trip to a separate Triton process is a dominant framework overhead~\cite{Villa2025ProgrammableAG}. We evaluate a four-stage backend progression that isolates each source of overhead in turn, culminating in an in-process Triton C-API path:
	\begin{enumerate}
		\item \textit{Direct PyTorch.} The actor is loaded from a training checkpoint and invoked in inference mode through the PyTorch Python API, with no export or inference server. Python and framework overhead dominate, so this stage serves as a development baseline.
		
		\item \textit{Triton gRPC.} The Triton Inference Server runs as a separate process and the dApp communicates with it over gRPC, matching the gRPC inference path of~\cite{Villa2025ProgrammableAG}; this retains Triton's model versioning, dynamic batching, and multi-backend support (ONNX Runtime, TensorRT) but pays the full Ops~5--8 round-trip cost.
		
		\item \textit{Triton in-process.} Triton is embedded directly inside the dApp process through its Python bindings, eliminating gRPC serialization and localhost networking; inference is invoked as a local function call with zero-copy DLPack tensor exchange.
		
		\item \textit{Triton C-API.} A pybind11 C++ extension drives the Triton Server C API directly, removing the remaining Python object churn on the per-slot inference path. Three optimizations target steady-state latency: (i)~the inference request object is built once at initialization and reused across all subsequent slots, (ii)~observation and output buffers are pre-allocated in CUDA pinned host memory to avoid host-side allocation on the hot path, and (iii)~a custom response allocator routes Triton's outputs into those pre-allocated buffers. The result is the lowest per-call overhead among the four backends.
	\end{enumerate}
	
	Section~\ref{subsec:latency_breakdown} reports a per-Op latency comparison across the four backends.
	
	\subsection{Multi-UE Batched Control}
	\label{subsec:multi-UE-batched-control}
	
	Practical deployments serve multiple UEs concurrently, yet neither the E3 reference dApps~\cite{Villa2025ProgrammableAG,Santhi2025,Olimpieri2025LibIQTR}, which target single-UE open-loop inference, nor existing RL-based link adaptation methods~\cite{Saxena:ITWC:2022,Ye:TVT:2023,Mota:GW:2019,An2024DRAGONAD}, which assume a single UE with a fixed PRB allocation, address the interaction between per-UE adaptation and a shared MAC scheduler. Extending the closed loop of Section~\ref{subsec:closed-loop-dApp} to $K$ concurrent UEs raises three architectural challenges.
	
	\textbf{C1: Per-UE state isolation.} A state manager indexed by Radio Network Temporary Identifier (RNTI) maintains for each UE its three most recent MCS indices, three most recent HARQ outcomes, and the running BLER estimate $\bar{\varepsilon}_t$ (Section~\ref{subsec:state_obs}) updated through an $O(1)$ circular buffer. The RNTI-to-row mapping is locked at initialization from the order of the first batch indication and preserved throughout the dApp's lifetime, guaranteeing strict alignment between the $i$-th row of the observation batch $\mathbf{O} \in \mathbb{R}^{K \times 13}$ and the $i$-th UE's state.
	
	\textbf{C2: MAC-RL interface isolation.} The production MAC scheduler executes three stages per slot: priority-weighted round-robin UE selection, PRB allocation, and OLLA-based MCS selection. The dApp invokes only the first two through a PRB-only scheduler variant, retrieves the per-UE PRB allocations, and applies its own MCS through the L2 E3 Agent override path of Section~\ref{subsec:closed-loop-dApp} (Step~6). Bypassing the OLLA stage prevents its internal offset from being updated on MCS values it never selected, which would otherwise corrupt any OLLA-only baseline reusing the same scheduler.
	
	\textbf{C3: Non-uniform per-UE resource dimensions.} When $N_\text{PRB} \bmod K \neq 0$, round-robin allocates $\lfloor N_\text{PRB}/K \rfloor + 1$ PRBs to a subset of UEs and $\lfloor N_\text{PRB}/K \rfloor$ to the rest, so per-UE H-matrix dimensions and Transport Block (TB) sizes differ. The per-UE subcarrier count is carried in each E3 indication for an independent reshape during feature extraction (Section~\ref{subsec:shared-memory}), and the exact per-UE TB sizes computed by the MAC scheduler are propagated through the E3 pipeline rather than re-derived from the average PRB count, eliminating a worst-case TB-size error of $1/\lfloor N_\text{PRB}/K \rfloor$ (the relative gap between adjacent per-UE PRB counts; ${\approx}\,2.9\%$ at $N_\text{PRB}=273$, $K=8$, and ${\approx}\,17\%$ at $N_\text{PRB}=52$, $K=8$).
	
	At each slot the dApp assembles $\mathbf{O}$, runs a batched forward pass emitting $K$ SINR offsets, and dispatches the per-UE MCS overrides as a batched E3 Control message; at training the $K$ UEs are batched as $K$ rollout streams under a shared policy, so the single-UE formulation of Section~\ref{sec:learned_olla} extends to multi-UE without algorithmic changes.
	
	\section{Simulation Results}
	\label{sec:sim_results}
	This section evaluates LOLLA against conventional OLLA, ILLA, and a Direct MCS-RL baseline representing prior RL-based link adaptation. We assess throughput, BLER constraint tracking, and end-to-end latency under 3GPP TDL channel models, covering Doppler sweeps, BLER target sweeps, cross-channel generalization, and multi-UE scaling.
	
	\subsection{Simulation Setup}
	\label{subsec:setup}
	Training and inference run on a single NVIDIA L20 GPU using the GPU-accelerated PHY simulator of Section~\ref{sec:system_arch} with 3GPP TDL channels~\cite{3GPP_TS38901}. The default configuration is in Table~\ref{tab:sim_params}, and per-experiment overrides are noted in the corresponding subsections.
	
	\textbf{Baselines.} LOLLA is compared against: (i)~\textit{conventional OLLA} with the standard step sizes $(\delta_\text{up}, \delta_\text{down}) = (0.1, 1.0)$\,dB yielding $\varepsilon_\text{eq} \approx 9.1\%$; (ii)~\textit{ILLA}, the LUT mapping with zero outer-loop offset; and (iii)~\textit{Direct MCS-RL}, a PPO agent with categorical action space $\mathcal{A} = \{0, 1, \ldots, 27\}$ that shares the same network, observations, and Lagrangian mechanism as LOLLA but bypasses the SINR-to-MCS LUT, mirroring prior RL-based link adaptation methods~\cite{Saxena:ITWC:2022,Ye:TVT:2023,Mota:GW:2019,An2024DRAGONAD} and isolating the contribution of the residual formulation. We additionally report fixed MCS indices $m \in \{5, 10, 15, 20\}$ as reference points.
	
	\textbf{Evaluation protocol.} Each trained policy is evaluated over $1{,}000$ slots per seed (deterministic action mean) starting from $\delta_0 = \Delta_0 = 0$. Results are averaged over 20 random seeds (mean~$\pm$~one standard deviation); minor deviations from Table~\ref{tab:sim_params} are noted inline.
	
	\begin{table}[t]
		\centering
		\caption{Default Simulation and Training Configuration}
		\label{tab:sim_params}
		\small
		\setlength{\tabcolsep}{4pt}
		\begin{tabular*}{\columnwidth}{@{\extracolsep{\fill}}lr}
			\toprule
			\textbf{Parameter} & \textbf{Value} \\
			\midrule
			\multicolumn{2}{l}{\textit{System and channel}} \\
			Antenna configuration / layers & $4 \times 2$ MIMO, $N_l = 2$ \\
			System bandwidth / $N_\text{PRB}$ & 100\,MHz / 273 PRBs \\
			Subcarrier spacing / center freq.\ & 30\,kHz / 3.5\,GHz \\
			MCS table & Table~5.1.3.1-1 (64QAM)~\cite{3GPP_TS38214} \\
			Channel model / delay spread & TDL-A / 100\,ns (nominal)~\cite{3GPP_TS38901} \\
			SNR range / change prob.\ & $[-5,\, 25]$\,dB / 0.3 per slot \\
			\midrule
			\multicolumn{2}{l}{\textit{PPO and network}} \\
			Learning rate $\alpha_0$ / floor $f_{\min}$ & $2.5{\times}10^{-4}$ / 0.2 \\
			Discount $\beta$ / GAE $\lambda_\text{GAE}$ & 0.99 / 0.95 \\
			Clip / entropy / value coeff.\ & 0.2 / 0.01 / 0.5 \\
			Rollout $T$ / minibatches / epochs & 128 / 4 / 4 \\
			MLP hidden dim.\ / layers & 64 / 2 (Tanh) \\
			Action range $\delta_{\max}$ & $\pm 31$\,dB \\
			\midrule
			\multicolumn{2}{l}{\textit{Lagrangian and training}} \\
			Initial / max dual $(\lambda_0, \lambda_{\max})$ & $(1.0, 50)$ \\
			Dual learning rate $\eta$ & 0.05 \\
			Default target BLER $\varepsilon_\text{target}$ & 0.091 \\
			Parallel envs.\ / total steps & 8 / $4{\times}10^6$ \\
			\bottomrule
		\end{tabular*}
	\end{table}
	
	\subsection{Training Convergence}
	\label{subsec:convergence}
	Fig.~\ref{fig:fig5_learning_curves} plots LOLLA and Direct MCS-RL training curves averaged over three seeds at $f_d = 100$\,Hz with $p_\text{SNR} = 0.3$, using $3 \times 10^6$ steps and a tighter Lagrangian range $(\lambda_0, \lambda_{\max}, \eta) = (1.0, 10, 0.08)$ but otherwise following Table~\ref{tab:sim_params}. Averaged over the last $10^5$ steps, both methods converge to $9.4\%$ BLER (within $0.3$\,pp of the $9.1\%$ target) and to comparable throughput, $252.4 \pm 2.6$\,Mbps for LOLLA versus $244.2 \pm 2.2$\,Mbps for Direct MCS-RL. LOLLA converges substantially faster in the early phase: at $200$K steps it already reaches $233.2 \pm 2.9$\,Mbps, whereas Direct MCS-RL lags at $206.0 \pm 2.2$\,Mbps, reflecting the strong initialization provided by the LUT prior. The Lagrangian dual variable $\lambda$ rises monotonically from $\lambda_0=1.0$ to approximately $4.6$ for LOLLA and $4.3$ for Direct MCS-RL, confirming the integral-controller behavior of Section~\ref{subsec:lagrangian_bler}; the narrow $\pm 1\sigma$ bands further indicate stable and reproducible training.
	
	\putFig{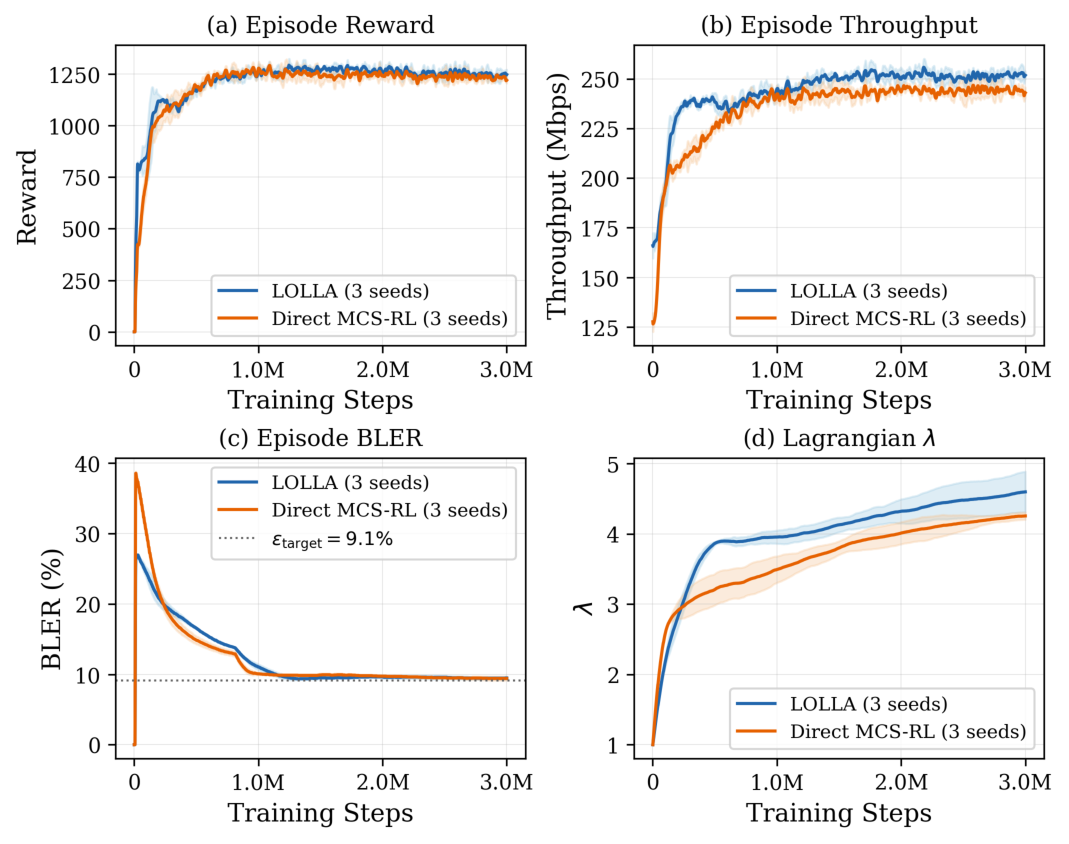}{Training convergence at $f_d = 100$\,Hz (TDL-A, $p_\text{SNR} = 0.3$). (a)~Episode reward. (b)~Episode throughput (Mbps). (c)~Episode BLER (\%) with $9.1\%$ target (dotted). (d)~Lagrangian $\lambda$. Mean $\pm\,1\sigma$ over 3 seeds (shaded). Blue: LOLLA; orange: Direct MCS-RL.}{3.45}
	
	\subsection{Throughput vs.\ Channel Dynamics}
	\label{subsec:throughput}
	We train PPO policies at five Doppler frequencies $f_d \in \{10,\, 50,\, 100,\, 200,\, 400\}$\,Hz (approximately $3$ to $120$\,km/h at $f_c = 3.5$\,GHz). At $f_d = 10$\,Hz the channel is nearly static and all four adaptive methods agree to within $3\%$ ($283$ to $290$\,Mbps), so the outer loop is essentially redundant. As $f_d$ grows, LOLLA delivers $284.0$, $250.2$, $179.4$, and $77.0$\,Mbps at $f_d = 50$, $100$, $200$, and $400$\,Hz, gains of $+15\%$, $+46\%$, $+92\%$, and $+75\%$ over OLLA, while holding the empirical BLER within $8.6\%$ to $9.3\%$. The gain peaks at $f_d = 200$\,Hz, where the coherence time $T_c \approx 1/(2 f_d) = 2.5$\,ms is shorter than OLLA's multi-slot convergence horizon yet LOLLA still adapts per slot from the channel features. Fig.~\ref{fig:fig6_doppler_sweep}(d) corroborates this: LOLLA's average MCS remains near $13$ for $f_d \leq 100$\,Hz, whereas OLLA collapses to $5.4$ and $1.9$ at $f_d = 200$ and $400$\,Hz as repeated NACK-triggered down-steps drag its offset negative. Direct MCS-RL stays within $\pm 3\%$ of LOLLA at every operating point; the residual formulation's advantage emerges only under distribution shift (Section~\ref{subsec:generalization}). The non-adaptive baselines collapse at high mobility, with ILLA exceeding $26\%$ BLER at $f_d \geq 100$\,Hz.
	
	\putFig{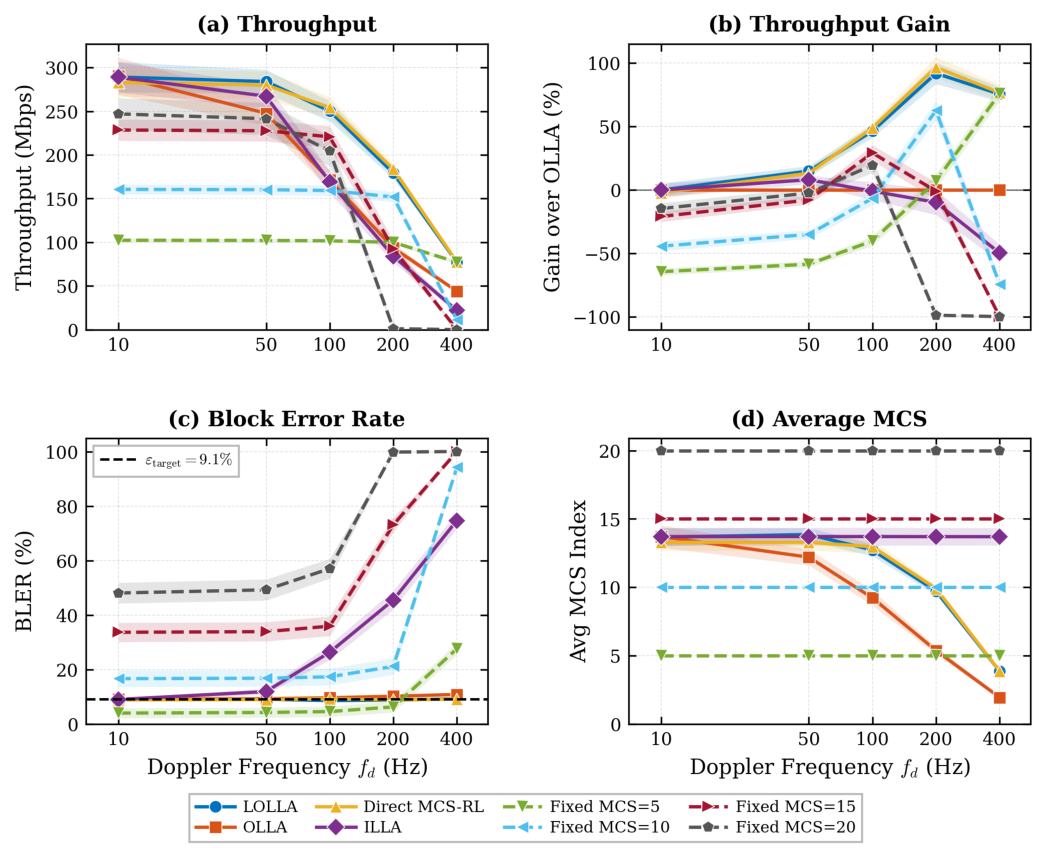}{Comparison of link adaptation methods across Doppler frequencies under TDL-A with $p_\text{SNR} = 0.3$. (a)~Per-UE throughput (Mbps). (b)~Throughput gain over OLLA (\%). (c)~Block error rate (\%) with the $\varepsilon_\text{target} = 9.1\%$ target (dashed). (d)~Average MCS index. Shaded regions denote $\pm\,1\sigma$ over 20 seeds.}{3.45}

	\subsection{Lagrangian BLER Constraint Control}
	\label{subsec:constraint_control}
	To validate the Lagrangian mechanism (Section~\ref{subsec:lagrangian_bler}), we sweep four BLER targets $\varepsilon_\text{target} \in \{1\%, 5\%, 9.1\%, 12\%\}$ at $f_d = 100$\,Hz, plus one unconstrained point obtained by setting the target so large that $\lambda$ converges to zero. For each constrained point, OLLA is configured with step sizes yielding the same equilibrium BLER.
	
	\textbf{Pareto frontier.} Fig.~\ref{fig:fig_lagrangian_combined}(a) shows that LOLLA dominates OLLA across the Pareto frontier: at every constrained operating point, LOLLA achieves substantially higher throughput at strictly lower BLER than OLLA. At the tightest target ($\varepsilon_\text{target} = 1\%$), OLLA overshoots the prescribed value to $2.6\%$ whereas LOLLA tracks $1.3\%$ precisely, confirming tight constraint satisfaction in the URLLC regime. At the unconstrained point, where both methods operate near $15.3\%$ BLER, the throughput ceiling gap reaches $50\%$, $272.6$\,Mbps for LOLLA versus $181.2$\,Mbps for OLLA. Direct MCS-RL attains comparable or slightly higher throughput but with larger BLER deviations from the prescribed targets (e.g., $5.9\%$ vs.\ the $5\%$ setpoint), consistent with the coarser granularity of discrete MCS selection.
	
	\textbf{Per-SNR BLER allocation.} A dedicated policy trained under fixed-SNR episodes ($\varepsilon_\text{target} = 5\%$) is evaluated at $13$ SNR points from $-5$ to $25$\,dB. Both methods achieve comparable per-SNR throughput [Fig.~\ref{fig:fig_lagrangian_combined}(b)], so BLER redistribution does not cost spectral efficiency. However, their BLER profiles differ [Fig.~\ref{fig:fig_lagrangian_combined}(c)]: OLLA holds a uniform $\sim 5\%$ across all SNR points, while LOLLA tightens BLER to $1\%$--$2\%$ across the $3$--$13$\,dB range and rises toward $\sim 10\%$ at $23$--$25$\,dB, in agreement with the flat-penalty analysis of Section~\ref{subsec:lagrangian_bler}.
	
	\putFig{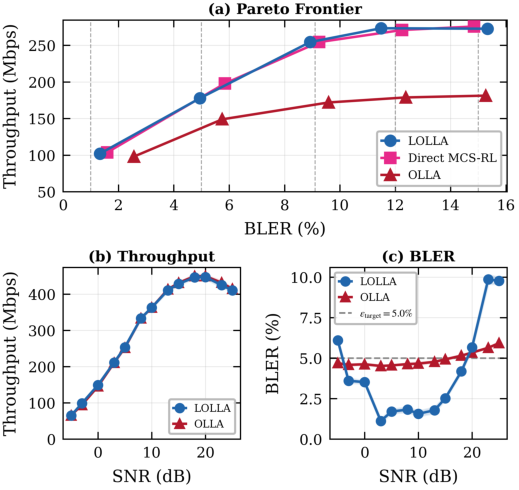}{Lagrangian BLER constraint analysis ($f_d = 100$\,Hz, TDL-A). (a)~Throughput versus BLER Pareto frontier ($p_\text{SNR} = 0.3$); dashed lines mark prescribed targets and the rightmost RL points correspond to the unconstrained ($\lambda = 0$) case. (b)~Per-SNR throughput and (c)~per-SNR BLER under fixed-SNR evaluation ($p_\text{SNR} = 0$, $\varepsilon_\text{target} = 5\%$); OLLA maintains a uniform $\sim 5\%$ BLER while LOLLA exhibits a non-uniform allocation.}{3.45}

	\subsection{Channel Model Generalization}
	\label{subsec:generalization}
	We evaluate zero-shot cross-channel generalization (Table~\ref{tab:channel_generalization}) by testing the TDL-A trained policy on three unseen channel models: TDL-B (NLOS, two dominant clusters), TDL-C (NLOS, extended multipath), and TDL-D (LOS, Rician fading), all at $f_d = 100$\,Hz. Both RL methods transfer well to the NLOS channels TDL-B and TDL-C, maintaining $45\%$ to $51\%$ throughput gains over OLLA with BLER between $8.8\%$ and $9.5\%$, within $1\%$ of the matched TDL-A performance. The LOS channel TDL-D, however, exposes a sharp robustness gap: LOLLA degrades to $27.0\%$ BLER but still delivers $165$\,Mbps ($+15\%$ over OLLA) because the SINR-to-MCS LUT prior bounds the offset to structurally valid MCS indices even when the policy is out-of-distribution. Direct MCS-RL, lacking this safety net, collapses to $52.7\%$ BLER and $112$\,Mbps ($-23\%$ versus OLLA), confirming that the residual formulation provides critical robustness under the Rayleigh-to-Rician distribution shift.
	
	\begin{table}[!htbp]
		\centering
		\caption{Cross-Channel Generalization (trained on TDL-A, $f_d = 100$\,Hz)}
		\label{tab:channel_generalization}
		\small
		\begin{tabular*}{\columnwidth}{@{\extracolsep{\fill}}lcccc}
			\toprule
			& \multicolumn{3}{c}{NLOS} & LOS \\
			\cmidrule(lr){2-4}\cmidrule(lr){5-5}
			\textbf{Method} & TDL-A$^\ast$ & TDL-B & TDL-C & TDL-D \\
			\midrule
			LOLLA    & 254\,/\,9.0  & 254\,/\,8.8  & 257\,/\,8.8  & 165\,/\,27.0 \\
			Direct MCS-RL   & 259\,/\,9.5  & 260\,/\,9.1  & 261\,/\,9.2  & 112\,/\,52.7 \\
			OLLA            & 171\,/\,9.6  & 175\,/\,9.6  & 177\,/\,9.6  & 144\,/\,9.7 \\
			\bottomrule
			\multicolumn{5}{l}{\footnotesize Throughput (Mbps)\,/\,BLER (\%). $^\ast$Matched (train = test).}
		\end{tabular*}
	\end{table}
	
	\subsection{Multi-UE Scaling}
	\label{subsec:multi-UEs}
	Using the multi-UE architecture of Section~\ref{subsec:multi-UE-batched-control}, we evaluate $K \in \{1, 2, 4, 8\}$ UEs sharing $N_\text{PRB} = 273$ PRBs under round-robin scheduling at $f_d = 100$\,Hz, with a separately trained policy per $K$ to match the per-UE PRB allocation ($\approx \lfloor 273/K \rfloor$ PRBs) and the resulting TB sizes.
	
	Fig.~\ref{fig:fig_multi_ue} shows that LOLLA delivers $254$--$258$\,Mbps for $K \leq 4$ and $229.9$\,Mbps at $K = 8$, with gains of $+27\%$ to $+48\%$ over OLLA. Aggregate throughput stays nearly flat for $K \leq 4$ and drops at $K = 8$, where the per-UE allocation shrinks to $34$ PRBs and the loss of frequency diversity together with the higher relative DMRS overhead reduce per-UE spectral efficiency for both schemes. The relative gain narrows monotonically as OLLA's conservative offset becomes less suboptimal in the narrowband regime, yet remains substantial at $K = 8$. Throughout the sweep, LOLLA holds the empirical BLER between $8.8\%$ and $9.1\%$, tracking the target tightly, while OLLA stays at $9.5\%$ to $9.6\%$.
	
	\putFig{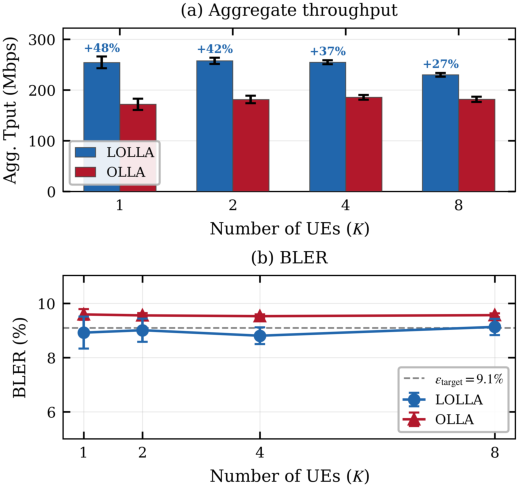}{Multi-UE scaling at $f_d = 100$\,Hz, $N_\text{PRB} = 273$, under round-robin scheduling for $K \in \{1,2,4,8\}$ UEs. (a)~Aggregate throughput with per-bar gain over OLLA annotated. (b)~BLER vs.\ number of UEs, with the $\varepsilon_\text{target} = 9.1\%$ target (dashed). Mean over 20 seeds.}{3.45}

	\subsection{End-to-End Latency Breakdown}
	\label{subsec:latency_breakdown}
	Table~\ref{tab:latency} decomposes the closed-loop latency into ten operations spanning the end-to-end data path (cf.~\cite[Table~1]{Villa2025ProgrammableAG}) for $K{=}1$ and $K{=}8$ UEs (C-API backend; medians over $N{=}1{,}000$ steps, NVIDIA L20). Data collection (Ops~1--4) takes $72/93$\,$\mu$s at $K{=}1/8$, backend-independent, with the increase reflecting the larger aggregate H-matrix transfer. On the dApp path (Ops~5--9), Triton input preparation (Op~6, parallelized across UEs via OpenMP at $K{=}8$) and CUDA model inference (Op~7) dominate at $124/102$ and $111/116$\,$\mu$s, together $86\%$ of dApp time, while gRPC serialization---the dominant overhead of the gRPC backend (Ops~5 and~8)---drops to $3$\,$\mu$s under the C API. Total dApp latency is $272/255$\,$\mu$s ($K{=}1/8$), a $56$--$59\%$ reduction from the gRPC backend ($623/615$\,$\mu$s).
	
	Including the control return (Op~10), the closed-loop end-to-end latency with the C API is $355/361$\,$\mu$s ($K{=}1/8$), within the $500$\,$\mu$s per-slot budget at $30$\,kHz SCS. Fig.~\ref{fig:fig_backend_latency} shows only the C API meets this budget; in-process is close (dApp $554/577$\,$\mu$s), while gRPC ($623/615$\,$\mu$s) and direct PyTorch ($1{,}975/2{,}852$\,$\mu$s) exceed it.
	
	\putFig{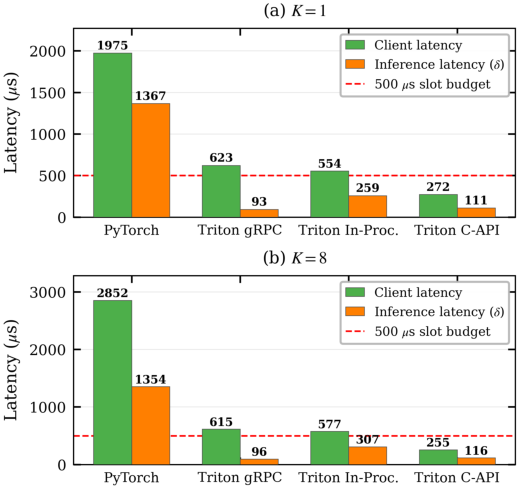}{Client latency (Ops~5--9) and inference latency ($\delta$, Op~7) across four backends for $K{=}1$ (top) and $K{=}8$ (bottom). Dashed line: 500\,$\mu$s slot budget (30\,kHz SCS).}{3.45}
	
	\begin{table}[t]
		\centering
		\caption{End-to-End Control-Loop Latency Decomposition ($\mu$s)}
		\label{tab:latency}
		\footnotesize
		\setlength{\tabcolsep}{3pt}
		\begin{tabular*}{\columnwidth}{@{\extracolsep{\fill}}clccc}
			\toprule
			\textbf{Op\textsuperscript{$\ast$}} & \textbf{Operation} & \textbf{Protocol} & \textbf{$K{=}1$} & \textbf{$K{=}8$} \\
			\midrule
			\multicolumn{5}{l}{\textit{Data collection}} \\
			1 & cuPHY copies data from GPU to CPU & memcpy & 35 & 45 \\
			2 & cuBB notifies ADL data is ready & Atomic & 18 & 22 \\
			3 & ADL copies data from CPU to SHM & memcpy & 18 & 25 \\
			4 & E3 Agent sends pointers to E3 Mgr & ZMQ & 1 & 1 \\
			& \textit{Subtotal (Ops~1--4)} & & \textit{72} & \textit{93} \\
			\midrule
			\multicolumn{5}{l}{\textit{dApp processing}} \\
			5 & Prepare input tensors (in-process)\textsuperscript{$\dagger$} & C API & 1 & 1 \\
			6 & Triton accesses and prepares input data & SHM & 124 & 102 \\
			7 & AI Model performs inference ($\delta$) & CUDA & 111 & 116 \\
			8 & Extract output tensors (in-process)\textsuperscript{$\dagger$} & C API & 2 & 3 \\
			9 & E3 Manager sends control to E3 Agent & ZMQ & 23 & 25 \\
			& \textit{Subtotal (Ops~5--9)} & & \textit{272} & \textit{255} \\
			\midrule
			\multicolumn{5}{l}{\textit{Control application}} \\
			10 & E3 Agent receives and applies control & API & 11 & 13 \\
			\midrule
			& \textbf{E2E total (C-API)} & & \textbf{355} & \textbf{361} \\
			\bottomrule
			\multicolumn{5}{l}{\scriptsize $^\ast$Op numbering follows the end-to-end data path; cf.~\cite[Table~1]{Villa2025ProgrammableAG}.} \\
			\multicolumn{5}{l}{\scriptsize $^\dagger$The C-API replaces gRPC serialization (Ops~5 and~8)} \\
			\multicolumn{5}{l}{\scriptsize ~~with in-process tensor preparation. The dApp Subtotal also} \\
			\multicolumn{5}{l}{\scriptsize ~~includes an $11/9\,\mu$s SINR-to-MCS LUT postprocess between Ops~8 and~9.} \\
			\multicolumn{5}{l}{\scriptsize Median over $N{=}1{,}000$ steps. NVIDIA L20 GPU, TDL-A, $f_d{=}100$\,Hz.}
		\end{tabular*}
	\end{table}
	
	\subsection{Ablation Studies}
	\label{subsec:ablation}
	
	We conduct two ablations at $f_d = 100$\,Hz, each removing one component of LOLLA [Table~\ref{tab:ablation}(a)]. Disabling dual ascent and fixing $\lambda = 0.25$ throughout training (A1) drives BLER to $11.7\%$, overshooting the $9.1\%$ target by $2.6$\,pp, whereas the full model tracks the target at $9.0\%$, confirming that adaptive $\lambda$ is necessary for tight constraint satisfaction. Restricting the observation to HARQ feedback only (A2, 7D: MCS history, ACK history, and running BLER) reduces throughput from $254$ to $185$\,Mbps, with BLER settling at $8.4\%$, $0.7$\,pp below the $9.1\%$ target. This undershoot reflects a partial-observability gap: without instantaneous channel and signal-quality features, the 7D policy cannot distinguish favorable from unfavorable slots and must hedge with a uniformly conservative offset. Since this 7D observation already subsumes OLLA's input, the remaining $+8\%$ over OLLA is attributable to the non-linear policy, while the additional $+37\%$ confirms that PHY-layer features are the dominant source of improvement.
	
	\begin{table}[!t]
		\centering
		\caption{Ablation Studies ($f_d = 100$\,Hz, TDL-A)}
		\label{tab:ablation}
		\small
		\setlength{\tabcolsep}{4pt}
		\textit{(a) Lagrangian and observation ($p_\text{SNR} = 0.3$)}\\[2pt]
		\begin{tabular*}{\columnwidth}{@{\extracolsep{\fill}}lcc}
			\toprule
			\textbf{Variant} & \textbf{Tput (Mbps)} & \textbf{BLER (\%)} \\
			\midrule
			LOLLA (13D)          & 254 & 9.0 \\
			(A1) Fixed $\lambda{=}0.25$ & 262 & 11.7 \\
			(A2) HARQ-only obs (7D)     & 185 & 8.4 \\
			OLLA                        & 171 & 9.6 \\
			\bottomrule
		\end{tabular*}
		\\[6pt]
		\textit{(b) H-matrix feature (mixed TDL-A/TDL-D)}\\[2pt]
		\begin{tabular*}{\columnwidth}{@{\extracolsep{\fill}}lccc}
			\toprule
			& \textbf{TDL-A} & \textbf{TDL-D} & \textbf{Mixed} \\
			\midrule
			LOLLA (13D) & 251\,/\,8.8  & 159\,/\,9.4  & 192\,/\,9.6 \\
			No $\mathbf{f}_\text{ch}$ (9D) & 216\,/\,7.2  & 164\,/\,10.4 & 181\,/\,9.3 \\
			OLLA          & 172\,/\,9.6  & 145\,/\,9.7  & 152\,/\,9.7 \\
			\bottomrule
			\multicolumn{4}{l}{\footnotesize Throughput (Mbps)\,/\,BLER (\%). 20 seeds, $f_d{=}100$\,Hz.}
		\end{tabular*}
	\end{table}
	
	To isolate the four H-matrix features $\mathbf{f}_\text{ch}$, we train a 13D and a 9D policy on mixed TDL-A/TDL-D episodes and evaluate each channel separately [Table~\ref{tab:ablation}(b)]. The 13D policy tracks the $9.1\%$ target on both channels, indicating that the H-matrix features let the agent identify the current fading profile and adapt accordingly. The 9D policy collapses to a compromise strategy that is overly conservative on TDL-A ($7.2\%$ BLER, $-35$\,Mbps) and overly aggressive on TDL-D ($10.4\%$); on mixed evaluation, 13D delivers $+6\%$ throughput at comparable BLER.
	
	\section{Conclusions}
	\label{sec:conclusions}
	We presented \textit{LOLLA}, the first closed-loop deep reinforcement learning controller for per-slot uplink link adaptation on a GPU-accelerated 5G NR stack. By learning a residual SINR offset that modulates the SINR-to-MCS lookup table and enforcing reliability through a Lagrangian dual variable, the agent provably matches or exceeds conventional OLLA under the idealized history-conditioned policy class while exploiting rich PHY/MAC telemetry inaccessible to its single-bit feedback. Realized as a dual-agent dApp on the E3 framework, LOLLA strictly dominates OLLA on the throughput-versus-BLER Pareto frontier, delivering $15\%$ to $92\%$ throughput gains over OLLA, scales to multiple concurrent UEs, and sustains sub-$500$\,$\mu$s end-to-end control latency, demonstrating that learned per-slot link adaptation is feasible on a production-grade GPU-accelerated RAN. Future work includes over-the-air validation with domain randomization for sim-to-real transfer~\cite{Ford2025Sim2FieldED}, joint MCS and MIMO-layer adaptation, multi-cell coordination, and extending the Lagrangian framework to other latency-critical control loops such as power control and beam management.
	
	\appendices
	
	\section{Proof of Proposition~\ref{prop:dominance}}
	\label{app:proof_dominance}
	We show that for every OLLA policy $\pi_\text{OLLA} \in \Pi_\text{OLLA}$, there exists a policy $\tilde{\pi} \in \Pi_\text{RL}$ that produces the same MCS at every time step on every sample path, and hence achieves the same expected return.
	
	\textit{Step 1: OLLA as a mapping.} A conventional OLLA policy $\pi_\text{OLLA} \in \Pi_\text{OLLA}$ is parameterized by step sizes $(\delta_\text{up}, \delta_\text{down})$ and generates offsets $\Delta_t$ via the recurrence
	\[
	\Delta_{t+1} = \Delta_t + \delta_\text{up}\,o_t - \delta_\text{down}\,(1 - o_t), \quad \Delta_0 = 0,
	\]
	where $o_t \in \{0,1\}$ is the HARQ outcome. Given any ACK/NACK history $\mathbf{o}_{0:t-1} = (o_0, \ldots, o_{t-1})$, the OLLA offset is a deterministic function of this history:
	\[
	\Delta_t = \delta_\text{up}\!\sum_{j=0}^{t-1} o_j \;-\; \delta_\text{down}\!\sum_{j=0}^{t-1}(1 - o_j).
	\]
	
	\textit{Step 2: Embedding in $\Pi_\text{RL}$.} The RL observation $\mathbf{o}_t \in \mathbb{R}^{13}$ includes the HARQ outcome $o_{t-1}$ as part of the ACK history $\mathbf{h}_\text{ack}$. Since each $\mathbf{o}_k$ contains $o_{k-1}$, the complete ACK history $(o_0, \ldots, o_{t-1})$ is recoverable from the observation sequence $(\mathbf{o}_1, \ldots, \mathbf{o}_t)$; hence the OLLA offset $\Delta_t$ from Step~1 is a deterministic, measurable function of $\mathbf{o}_{1:t}$. The embedding policy computes $\Delta_t$ from $\mathbf{o}_{1:t}$ and outputs $\delta_t = \mathrm{clip}(\Delta_t,\, -\delta_\text{max},\, \delta_\text{max})$. If $|\Delta_t| \le \delta_\text{max}$, the MCS is identical to OLLA. Otherwise, LUT saturation guarantees the same MCS. Let $\tilde{\gamma}_t \triangleq \min(\gamma_t, \gamma_\text{max}) \in [\gamma_\text{min},\, \gamma_\text{max}]$. When $\Delta_t > \delta_\text{max}$: $\tilde{\gamma}_t + \delta_\text{max} \ge \gamma_\text{min} + (\gamma_\text{max} - \gamma_\text{min}) = \gamma_\text{max}$; since $\gamma_\text{max}$ exceeds all MCS thresholds by design, both $\tilde{\gamma}_t + \delta_\text{max}$ and $\tilde{\gamma}_t + \Delta_t \;(> \tilde{\gamma}_t + \delta_\text{max})$ yield the maximum MCS. When $\Delta_t < -\delta_\text{max}$: $\tilde{\gamma}_t + \Delta_t < \tilde{\gamma}_t - \delta_\text{max} \le \gamma_\text{max} - \delta_\text{max} \le \gamma_\text{min} \le \gamma_\text{LUT}^{\min}$, so both yield MCS\,$= 0$. In every case the MCS is identical; since the same MCS produces the same HARQ feedback under identical channel conditions, $\Delta_{t+1}$ evolves identically by Step~1. By induction the entire trajectory matches on every sample path, and the embedding policy belongs to $\Pi_\text{RL}$ for any step-size pair $(\delta_\text{up}, \delta_\text{down})$. Our default $\delta_\text{max} = 31$\,dB is the smallest integer value satisfying this condition for the LUT bounds ($\gamma_\text{max} = 25.99$\,dB, $\gamma_\text{LUT}^{\min} = -4.57$\,dB).
	
	\textit{Step 3: Performance bound.} Since every OLLA policy has a trajectory-equivalent representative in $\Pi_\text{RL}$, the RL optimum is at least as large:
	\[
	\sup_{\pi \in \Pi_\text{RL}} (1-\beta)\,\mathbb{E}_\pi\!\left[\sum_{t=0}^{\infty} \beta^t \bar{r}_t\right] \ge \sup_{\pi \in \Pi_\text{OLLA}} (1-\beta)\,\mathbb{E}_\pi\!\left[\sum_{t=0}^{\infty} \beta^t \bar{r}_t\right].
	\]
	Informally, the inequality is strict whenever there exists a state reachable under OLLA where the richer observation $\mathbf{o}_t$ (channel features, signal quality, MCS history) enables a strictly better action than the one prescribed by the OLLA update rule conditioned only on $o_{t-1}$. This is the generic case in time-varying fading channels, where per-subcarrier channel estimates provide information about upcoming SNR transitions that binary HARQ feedback cannot capture.
	
	\section{Proof of Proposition~\ref{prop:duality}}
	\label{app:proof_duality}
	We prove each part separately.
	
	\textit{Part (i): Strong duality.} Both the objective~(\ref{eq:cmdp_obj}) and the BLER constraint~(\ref{eq:cmdp_bler}) are expressed in discounted form. Following Altman's framework~\cite[\S2.2, eq.~(2.4)]{Altman1999CMDP}, which uses the $(1-\beta)$-normalized discounted cost by convention, the finite CMDP is equivalent to a linear program (LP) over discounted occupation measures $\mu(s,a) \triangleq (1-\beta)\sum_{t \ge 0} \beta^t \Pr(s_t{=}s, a_t{=}a) \ge 0$~\cite[Theorem~3.3]{Altman1999CMDP}, with flow-conservation constraints~\cite[eq.~(3.5)]{Altman1999CMDP} and the BLER constraint $\sum_{s,a}\mu(s,a)\,c(s,a) \le \varepsilon_\text{target}$. The feasible set is non-empty (any policy selecting the lowest MCS achieves $J_c(\pi) \approx 0 < \varepsilon_\text{target}$). The set of occupation measures $\mathbf{L}^\beta$ is convex and compact, and coincides with the set attained by stationary policies~\cite[Theorem~3.2]{Altman1999CMDP}. Since $\mathbf{L}^\beta$ is convex-compact, $\mathcal{L}$ is continuous in $(\pi,\lambda)$, and $\lambda\ge 0$ is restricted to a bounded interval by the dual clip, the minimax theorem yields strong duality with attained optima: $\max_\pi \min_{\lambda \ge 0} \mathcal{L} = \min_{\lambda \ge 0} \max_\pi \mathcal{L}$~\cite[Theorem~3.6(i),(iii)]{Altman1999CMDP}. Equivalently, the primal LP over occupation measures~\cite[\S3.1]{Altman1999CMDP} and its dual over value functions and Lagrange multipliers~\cite[\S3.4]{Altman1999CMDP} attain the same optimal value~\cite[Theorem~3.7]{Altman1999CMDP}.
	
	\textit{Part (ii): Complementary slackness.} The primal LP has a finite optimum and the feasible set is non-empty, so by LP complementary slackness~\cite[\S5.5.2]{Boyd2004} the primal-dual optima $(\mu^*, \lambda^*)$ satisfy $\lambda^*(J_c(\pi^*) - \varepsilon_\text{target}) = 0$. Equivalently, under Slater's condition (which holds by construction), Theorem~3.6(iii) of~\cite{Altman1999CMDP} guarantees that Lagrange multipliers exist and satisfy the Karush--Kuhn--Tucker (KKT) conditions, including $\lambda^* \ge 0$ and $\lambda^*(J_c(\pi^*) - \varepsilon_\text{target}) = 0$.
	
	\textit{Part (iii): Function approximation.} Slater's condition holds by construction (the lowest-MCS policy is strictly feasible). If $\Pi_\Theta$ is $\varepsilon$-universal~\cite[Definition~1]{Paternain:NIPS:2019} (i.e., for every policy $\pi$ there exists $\theta$ with $\max_{s\in\mathcal{S}}\int_{\mathcal{A}}|\pi(a|s) - \pi_\theta(a|s)|\,da \le \varepsilon$), then Theorem~2 of~\cite{Paternain:NIPS:2019} establishes that the parametric duality gap, bounding both optimality loss and constraint violation, is $O(\varepsilon/(1-\beta))$. As network capacity grows, $\varepsilon \rightarrow 0$.
	
	\bibliography{thebib}
	
\end{document}